# Unifying Modular and Core-Periphery Structure in Functional Brain Networks over Development

Shi Gu[a,b,c], Cedric Huchuan Xia[b], Rastko Ciric[b], Tyler M. Moore[b], Ruben C. Gur[b], Raquel E. Gur[b], Theodore D. Satterthwaite[b,+], Danielle S. Bassett[c,d,e,f,+,*]

[a] School of Computer Science and Engineering, University of Electronic Science and Technology of China, Chengdu, China

[b] Department of Psychiatry, University of Pennsylvania, Philadelphia, PA 19104, USA

[c] Department of Bioengineering, School of Engineering and Applied Science, University of Pennsylvania, Philadelphia, PA

[d] Department of Physics and Astronomy, College of Arts and Sciences, University of Pennsylvania, Philadelphia, PA 19104, USA

[e] Department of Electrical and Systems Engineering, University of Pennsylvania, Philadelphia, PA 19104, USA

[f] Department of Neurology, Perelman School of Medicine, University of Pennsylvania, Philadelphia, PA 19104, USA

[+] Contributed Equally to this work.
[*] To whom correspondence should be addressed: dsb@seas.upenn.edu.

**Corresponding Author:**

Danielle S. Bassett
Office: 113 Hayden Hall
T: 215-746-1754
F: 215-573-2071
Email: dsb@seas.upenn.edu
Mail:210 S. 33rd Street
240 Skirkanich Hall
Philadelphia, PA 19104-6321

**Abstract**

At rest, human brain functional networks display striking modular architecture in which coherent clusters of brain regions are activated. The modular account of brain function is pervasive, reliable, and reproducible. Yet, a complementary perspective posits a core-periphery or *rich-club* account of brain function, where hubs are densely interconnected with one another, allowing for integrative processing. Unifying these two perspectives has remained difficult due to the fact that the methodological tools to identify modules are entirely distinct from the methodological tools to identify core-periphery structure. Here we leverage a recently-developed model-based approach -- the weighted stochastic block model -- that simultaneously uncovers modular and core-periphery structure, and we apply it to fMRI data acquired at rest in 872 youth of the Philadelphia Neurodevelopmental Cohort. We demonstrate that functional brain networks display rich meso-scale organization beyond that sought by modularity maximization techniques. Moreover, we show that this meso-scale organization changes appreciably over the course of neurodevelopment, and that individual differences in this organization predict individual differences in cognition more accurately than module organization alone. Broadly, our study provides a unified assessment of modular and core-periphery structure in functional brain networks, providing novel insights into their development and implications for behavior.

**Introduction**

Human cognition and behavior are grounded in the brain's complex neuroanatomical architecture and reflected in its functional dynamics. Recent efforts in *network neuroscience* (Bassett and Sporns 2017), an interdisciplinary fusion of network science and neuroimaging, have begun to uncover network-level explanations for this architecture (Bullmore and Sporns 2009, 2012) and mechanisms for these dynamics (Hutchison et al. 2013; Chaudhuri et al. 2015). Here, network nodes are defined as brain regions and network edges are defined as summary statistics, reflecting either interregional tractography in a structural brain graph, or statistical relationships among regional activity time series in a functional brain graph (Fornito et al. 2013). The formal representation of the brain as a graph facilitates the application of graph theoretical tools to link the graph's topology and dynamic properties to higher-order cognition (Bassett et al. 2011, 2015; Chai et al. 2016), including changes in cognitive capabilities that accompany development (Gu et al. 2015; Chai et al. 2017)}.

In the context of human neuroimaging, one particularly important set of tools – community detection – offers methods to decompose a network into modules or communities (Sporns and Betzel 2016). A common example applied to both structural (Sporns et al. 2005; Hagmann et al. 2007) and functional (Van Den Heuvel and Pol 2010) brain graphs is modularity maximization (Newman 2006), which identifies groups of nodes such that nodes within a group are more densely connected to other nodes in their group than anticipated in an appropriate random network null model. While useful, modularity maximization and similar approaches such as *Infomap* (Rosvall and Bergstrom 2008) make the important assumption that the brain's mesoscale architecture is best characterized by modules that are maximally independent from one another. Such an assumption is not without support from philosophical work in neuroscience and psychology over the last few decades. For example, the Fodorian view is that modules are characterized by informational encapsulation, with little need to refer to other psychological systems in order to operate (Fodor 1983).

Nevertheless, despite its historical roots, recent empirical evidence and emerging theoretical understanding has begun to call into question the notion that the brain network architecture supporting complex cognition is best characterized by largely independent modules. At a neuroanatomical level, the pattern of white matter connections displays structural connectivity among modules (Hagmann et al. 2008), and the strength of that inter-modular connectivity differs according to the modules involved (Betzel et al. 2017, 2019). The heterogeneous pattern of strong and weak inter-modular connectivity at the large-scale level of white matter structure is thought to facilitate and constrain integration of neural activity across diverse cognitive systems, enabling their collective function (Baum et al. 2017). Consistent with these observations of underlying structure, at the physiological level, the pattern of functional connections also displays non-trivial integration between modules, with some module pairs being more or less integrated than others (Meunier et al. 2009). For example, executive modules such as the fronto-parietal system tend to be more integrated with other brain systems (Power et al. 2013). The strength of between-module connectivity changes over development (Gu et al. 2015), differs in individuals in accordance with cognitive capabilities (Satterthwaite, Wolf, et al. 2015), and is altered in psychiatric disease in both adults (Sharma et al. 2017) and youth (Satterthwaite, Vandekar, et al. 2015), underscoring its relevance to brain function.

An important open question is whether there is a simple organizing principle that explains the heterogeneous patterns of inter-module connectivity observed in both anatomy and function. For example, are modules connected in a small-world organization, where modules tend to form clusters enabling local integration between modules, with a few modules extending topologically long-distance connections to other modules enabling global integration? Or perhaps a few modules serve as hubs in the inter-module network, while most modules are sparingly connected. While the literature has not settled on conclusive answers to these questions, one coarse-grained topological principle that has been shown to account for some of this heterogeneity is rich-club organization (Colizza et al. 2006), a specific sort of core-periphery structure (Borgatti and Everett 2000; Rombach et al. 2014; Zhang et al. 2015) whereby a set of highly connected and strongly interconnected hubs in the brain is complemented by a more sparsely connected network periphery (van den Heuvel and Sporns 2013). The core-periphery structure of underlying anatomy has important implications for the dynamics that can occur atop them (Betzel et al. 2016), to some degree explaining the core-periphery organization that is also observed in functional networks estimated from fMRI data collected during task performance (Ekman et al. 2012; Bassett, Wymbs, et al. 2013) and during the resting state (Gu et al. 2017).

The observation that both modular structure and core-periphery structure characterize brain graphs raises several challenging questions. How are modules related to cores, or to peripheries? Is there a simple organizational principle explaining these two characteristics of network architecture? How is that principle altered across development or manifested in different individuals? Answering these questions is particularly challenging because the methodological tools to identify modules are entirely distinct from the methodological tools to identify core-periphery structure. Here we employ a recently-developed model-based approach -- the weighted stochastic block model (WSBM) (Aicher et al. 2014) -- that simultaneously uncovers modular structure, core-periphery structure, and other organizational features that can occur in networks with richly and heterogeneously connected modules (Betzel et al. 2017, 2019). We apply the WSBM to functional networks extracted from resting state data acquired in a large sample of youth imaged as part of the Philadelphia Neurodevelopmental Cohort (Satterthwaite et al. 2014). We hypothesized that functional brain networks would display rich meso-scale organization beyond that sought by modularity maximization techniques, that richer meso-scale organization would change over the course of normative neurodevelopment, and that individual differences in this organization would be more predictive of individual differences in cognition than individual differences in module organization alone (Gordon et al. 2016).

**Materials and Methods**

**Participants.** Resting state functional magnetic resonance imaging (fMRI) data were obtained from $n = 1601$ youth who participated in a large community-based study of brain development, now known as the Philadelphia Neurodevelopmental Cohort (PNC) (Satterthwaite et al. 2014). The present sample includes $n = 872$ participants between the ages of 8 and 22 years (mean $= 15.65$, s.d. $= 3.3$, 377 males, 495 females). A total of 729 of the initial 1601 participants were excluded for the following reasons: medical problems that may impact brain function, incidental radiologic abnormalities in brain structure, a poor quality T1 image (Rosen et al. 2018), high motion during the resting fMRI scan (see below), or poor coverage in task-free BOLD images.

**Neurocognitive battery.** Cognition was measured outside of the scanner using the Penn Computerized Neurocognitive Battery (CNB) (Gur et al. 2010, 2012). Briefly, the 1-hour CNB was administered to all participants, and consisted of 14 tests that evaluated a broad range of cognitive functions. Twelve of the tests measure both accuracy and speed, while two of the tests (motor and sensorimotor) measure only speed. Here, we used the factor score for executive efficiency from a best-fitting four-factor solution comprising tests from the executive function domain, attention, abstraction, and working memory (Moore et al. 2015a). The tests contributing to the executive efficiency score include the Penn Continuous Performance Test, the Letter N-Back task, and the Penn Verbal Reasoning Test (Moore et al. 2015b, 2016, 2017). In the present study, we use this factor score for executive efficiency as our primary measure, hereafter referred to as simply *executive function*.

**Imaging data acquisition and preprocessing.** MRI data were acquired on a 3 Tesla Siemens Tim Trio whole-body scanner and 32-channel head coil at the Hospital of the University of Pennsylvania. A T1-weighted image was acquired for each subject. All subjects underwent functional imaging (TR = 3000 ms; TE = 32 ms; flip angle = 90 degrees; FOV = 192 $\times$ 192 mm; matrix = 64 $\times$ 64; slices = 46; slice thickness = 3 mm; slice gap = 0 mm; effective voxel resolution = 3.0 $\times$ 3.0 $\times$ 3.0 mm) during a 6-minute resting-state sequence during which a cross-hair for fixation was displayed.

Raw resting-state fMRI BOLD data was processed using a top-performing preprocessing pipeline that has been shown to markedly reduce the impact of in-scanner motion (Satterthwaite et al. 2013; Ciric et al. 2017). This pipeline included: 1) distortion correction with FSL's FUGUE utility, 2) template registration with MCFLIRT, 3) de-spiking with AFNI's 3DDESPIKE utility, 4) demeaning to remove linear or quadratic trends, 5) boundary-based registration to individual high-resolution structural image, 6) 36-parameter global confound regression, and 7) first-order Butterworth filtering to retain signal in the 0.01 to 0.08 Hz range. For all analyses of fMRI data, we excluded subjects with incomplete data or excessive head motion (mean relative displacement > 0.5mm or maximum displacement > 6 mm).

**Network construction.** Here we model the resting state functional connectivity of each subject as a network (Bassett et al. 2018). We begin by noting that a simple networked system can be represented by the graph $\mathcal{G} = (\mathcal{V}, \mathcal{E})$, where $\mathcal{V}$ and $\mathcal{E}$ are the vertex and edge sets, respectively. Let $a_{ij}$ be the weight associated with the edge $(i,j) \in \mathcal{E}$, and define the *weighted adjacency matrix* of $\mathcal{G}$ as $A = a_{ij}$, where $a_{ij} = 0$ whenever $(i,j) \notin \mathcal{E}$. In this study, each network node represents one of 333 cortical areas specified by the Gordon atlas (Fig. 1A). Each network edge was defined as the Pearson correlation coefficient between the regional mean BOLD time series of region i and region j, followed by the application of a Fisher's r-to-z-transformation (Fig. 1B).

**Weighted stochastic block model.** Fundamentally, the stochastic block model is a generative model for random graphs that seeks to partition nodes into sets such that nodes with similar patterns of binary connectivity to the rest of the brain are grouped together with one another. Here we applied a recent extension of this method to weighted graphs, commonly referred to as the weighted stochastic block model (Aicher et al. 2014). Formally, we follow the notation in (Aicher et al. 2014), which describes the generative model for weighted pairwise interactions among n vertices, with an exponential family distribution $\mathcal{F}$ and a block structure $\mathcal{R}$. For a subject s with an adjacency matrix $\mathbf{A}^s$, the probability of the graph is

$$Pr(\mathbf{A}^s|\mathbf{z}^s, \boldsymbol{\theta}^s, \mathcal{M}_{\mathcal{F},\mathcal{R}}^s) = \prod_{i \leq j} f^s\left(A_{ij}^s \Big| \theta_{\mathcal{R}(z_i, z_j)}^s\right), \quad (1)$$

where $\mathbf{z}^s$ are the community labels, and $\boldsymbol{\theta}^s$ is the matrix of edge bundle parameters $\boldsymbol{\theta}$. For an estimated model $\mathcal{M}_{\mathcal{F},\mathcal{R}}$, the log-evidence score

$$\mathcal{L}_{\mathcal{M}_{\mathcal{F},\mathcal{R}}} = \log Pr(\mathbf{A}|\mathcal{M}_{\mathcal{F},\mathcal{R}}), \quad (2)$$

is used to quantify the goodness of fit and to inform model selection.

Next, we define summary statistics that can be used to describe the organization of the estimated block structure, and we also consider how to extend the model to reflect the shared structure in a group of subjects. We begin by supposing that we have N subjects with adjacency matrix $\mathbf{A}^1, \ldots, \mathbf{A}^N$ and the optimized density $f^s$ with $Pr(\mathbf{A}^s) = f^s(\mathbf{A}^s)$. Because the WSBM groups nodes together into sets (or blocks), we have the opportunity to quantify the inter-block connectivity strength as well as the intra-block connectivity strength. For the block average adjacency matrix $\Omega^s = \omega_{mn}^s$ with corresponding block set $C_m$, the block average strength between block m and n of subject s is defined as

$$\omega_{mn}^s = \begin{cases} \dfrac{\sum_{i \in C_m, j \in C_n, i \neq j} A_{ij}^s}{|C_m| \cdot |C_m| - |C_m|} & \text{if } m = n, \\ \dfrac{\sum_{i \in C_m, j \in C_n} A_{ij}^s}{|C_m| \cdot |C_n|} & \text{if } m \neq n, \end{cases} \quad (3)$$

and the block allegiance matrix $\mathbf{F}^s$ representing the expected strength between two regions for each subject is defined as

$$F_{ij}^s = \mathbb{E}^s\left(A_{ij}^t \Big| \theta_{\mathcal{R}(z_i, z_j)}^s\right) = \omega_{z_i, z_j}^s, \quad (4)$$

where $\mathbb{E}^s$ is the expectation of the estimated model of the s-th subject and $z_i, z_j$ are the block labels of region i and region j.

The mean block average strength $\overline{\boldsymbol{\Omega}} = \{\overline{\omega}_{mn}\}$ is then defined as the mean of $\Omega^s$ across subjects:

$$\bar{\omega}_{mn} = \frac{1}{N}\sum_{s=1}^{N} \omega_{mn}^{s}, \tag{5}$$

and the average block allegiance matrix is defined as the mean of $F_{ij}^{s}$ across subjects:

$$F_{ij} = \frac{1}{N}\sum_{s=1}^{N} F_{ij}^{s}, \tag{6}$$

where $N$ is the number of subjects.

**Identification of Group-level Hierarchical Meso- and Macro- Scale Structures.** In order to obtain an estimated hierarchical block structure at the group level, we designed a four-step procedure comprised of repeated application of the WSBM. At the first scale of the hierarchy (Step 1), we applied the WSBM-based clustering to single-subject functional connectivity matrices, and we varied the number of blocks $k_s$ from $k_{min}$ to $k_{max}$ in increments of one to obtain individual-level estimates of block structure. To move to the next level of the hierarchy (Step 2), we computed the average block allegiance matrix $F$ for each $k_s$ using Eq. 6. We examined the effect of $k_s$ on $F_{ij}$ to determine the value of $k_s$ above which $F_{ij}$ remained relatively stable. For the remaining steps in this hierarchical procedure, we fixed $F$ at this stable architecture. Next (Step 3), we applied the WSBM to the fixed $F$ and tuned the number of blocks $k$ to attain the highest log-evidence. This procedure produced an optimal group-level meso-scale structure. From this structure, we next computed the associated meso-scale connectivity matrix, where each node represented a block from the WSBM and where each edge was weighted by the mean connectivity strength between pairs of blocks. Finally (Step 4), we applied the WSBM to this meso-scale connectivity matrix to obtain a macro-scale structure that represented the relationships among blocks. Through this four-step procedure, we acquired the individual block structures as well as representative group-level meso- and macro- scale structures, which we study throughout the article.

**Quantitatively examination of core-periphery structure: estimating the regional core-score.** We used the core scores defined in (Rombach et al. 2014) to perform our core-periphery analysis. The core score for node i is defined as

$$CS(i) = Z \sum_\gamma C_i(\gamma) \times R_\gamma, \tag{7}$$

where Z is the normalization factor so that $\max_i[CS(i)] = 1$, $\gamma = (\alpha, \beta)$ is the scale parameter for local core value

$$C_i(\gamma) = C_i(\alpha, \beta) = \frac{1}{1+\exp(-(i-|\mathcal{V}|\beta)\times \tan(\pi\alpha/2))}, \tag{8}$$

Where $|\mathcal{V}|$ is the number of nodes, and $R_{ij} = \sum_{ij} A_{ij} C_i C_j$ is the core quality.

**Identification of Inter-Block Relations in the Meso-scale Structure.** After examining the existence of core-periphery relationships in the resting functional brain networks, we further investigated how different blocks interact with each other on the meso-scale. Consider a given pair of blocks m, n with associated within-block mean strengths $\omega_{mm}$ and $\omega_{nn}$, and with associated between-block mean strength $\omega_{mn}$. We say that this pair forms a *core-periphery pair* when $|\omega_{mm}| \leq |\omega_{mn}| \leq |\omega_{nn}|$ (or $|\omega_{nn}| \leq |\omega_{mn}| \leq |\omega_{mm}|$), a *bipartite pair* when $|\omega_{mn}| \geq \max\{|\omega_{mm}|, |\omega_{nn}|\}$, or an *independent pair* when $|\omega_{mn}| \leq \min\{|\omega_{mm}|, |\omega_{nn}|\}$.

After identifying the core-periphery pairs in the functional connectivity matrix, it is necessary to ask whether the findings are statistically significant. Here we conduct a non-parametric permutation test to assess statistical significance. For a subject s with a block average adjacency matrix $\Omega^s$, we define $H^s = \eta_{mn}^s$ as its characteristic matrix of core-periphery structure, where

$$\eta_{mn}^s = \begin{cases} -1 & \text{if } \omega_{mm} < \omega_{mn} \leq \omega_{nn} \\ 1 & \text{if } \omega_{nn} \leq \omega_{mn} < \omega_{mm} \\ 0 & otherwise \end{cases}, \qquad (9)$$

The average core-periphery role $H$ is then defined as the mean of $H^s$ across subjects:

$$H = \frac{1}{N} \sum_{s=1}^{N} H^s. \qquad (10)$$

To determine whether the core-periphery pairs are located uniformly at random, we consider a null model in which core-periphery pairs are chosen uniformly at random for each subject. Under the null hypothesis instantiated by this model, the probability $\eta_{mn}^0$ that a pair $mn$ is a core-periphery pair is estimated as $(\sum_{s=1}^{N} |\eta_{mn}^s|)/N$ and its standard deviation is $\sigma_{mn}^0 = \sqrt{\eta_{mn}^0(1-\eta_{mn}^0)/N}$, where $N$ is the number of subjects. The z-score of a pair $mn$ being a core-periphery pair is then defined as

$$z_{mn} = \frac{|\eta_{mn}| - \eta_{mn}^0}{\sigma_{mn}^0}, \qquad (11)$$

which represents the regularized difference between the frequency of pair $mn$ and the expected probability of a random pair to appear as a core-periphery interaction. The associated p-values are denoted as $p_{mn}$ and calculated following the standard normal distribution. By applying an FDR correction for multiple comparisons, we can binarize the average core-periphery role $H$ to $\widetilde{H} = \{\widetilde{\eta_{mn}}\}$, where $\widetilde{\eta_{mn}} = 1$ for $p_{mn}^{corrected} < p_{FDR}$, representing the existence of a significant core-periphery pair.

**Identification of Core-periphery Junctions.** After identifying core-periphery interactions, we next explored the possibility of complex conjunctions among multiple blocks. Specifically, by integrating core-periphery pairs with common blocks, we recognized a pattern which we call a *core-periphery junction*. Mathematically, a core-periphery junction is defined as a connected component of the core-periphery relationship graph (see Eq. 9~10). Intuitively, a junction represents a sector of the system in which specific blocks act as linkers between a core block and a periphery block. By examining this integrated structure, we can probe relations between more integrative structures (cores) and more segregated structures (peripheries).

**Modularity Maximization.** The modular structure is obtained by modularity maximization with a Newman-Girvan null model, where the modularity function is defined as

$$Q = \sum_{ij} \left[ \left( A_{ij}^+ - \gamma^+ \frac{p_i^+ p_j^+}{2\mu^+} \right) + \left( A_{ij}^- - \gamma^- \frac{p_i^- p_j^-}{2\mu^-} \right) \right] \delta(c_i, c_j) \ , \quad (12)$$

where $A^+$ is the positive part of the weighted adjacency matrix A, $A^-$ is the negative part of A, $p_i^+ = \sum_j A_{ij}^+$, $p_i^- = \sum_j A_{ij}^-$, $\mu^+ = \sum_{ij} A_{ij}^+$, $\mu^- = \sum_{ij} A_{ij}^-$, $c_i$ is the community assignment for node i, and $\gamma^+$, $\gamma^-$ are resolution parameters that tune the relative size of modules detected. We optimized this modularity quality function using a Louvain-like locally greedy algorithm (Jutla et al. n.d.; Blondel et al. 2008). We performed a parameter sweep across values of $\gamma$ (Bassett, Porter, et al. 2013) to identify a $\gamma$ value that produced the same number of modules as the number of blocks in the WSBM. At that $\gamma$ value we then identified a consensus partition across multiple runs of the algorithm.

**Assessing the variability or consistency of intra-block edge weights.** We assessed the performance of the WSBM and modularity maximization algorithms by calculating the variance in the edge weights located within blocks or modules. Intuitively, this metric served to probe the fit of the model to the data, with low edge weight variability indicating greater internal homogeneity of blocks or modules. Specifically, for subject s with $k_s$ blocks or modules in its adjacency matrix, we denote the standard deviation of edge weights within block m's upper triangle as $\sigma_m^s$. Then, we calculate the standard deviation $\overline{\sigma^s}$ as the squared mean of $\sigma_m^s$'s, i.e.

$$\overline{\sigma^s} = \sqrt{\frac{\sum_{m=1}^{k_s}(\sigma_m^s)^2}{k_s}}, \tag{13}$$

**Statistical Testing.** Throughout the majority of the results section, we reported standard parametric statistical tests and associated p-values. In the context of Fig. 4, 5, we employed non-parametric permutation testing due to the non-normal distribution of the data. First, the p-value reported in Fig. 4 was computed via a permutation test in which each of the 100,000 random instantiations were created by randomly shuffling the block labels and recalculating the within-block mean of core-scores. These estimates then formed the null distribution of block-averaged core scores shown in the bottom row in Fig. 4. Second, the p-value reported in Fig. 5B was computed via a two-sample t-test between the distribution of the number of core-periphery pairs in the partition by WSBM and that in the partition obtained in a non-parametric permutation-based null model. For each subject, we randomly shuffled the block association achieved by WSBM for each node with the block size retained. Next, we applied the rule implemented in Fig. 5A and Eq. 9 to identify the number of core-periphery pairs for each subject. These estimates formed the null distribution that we used in the two-sample t-test.

In the context of Fig. 6, the correlations were computed between the block connectivity strength and age, after partialing out the effects of mean framewise displacement and sex. The standard p-values associated with partial correlations were reported here. In the context of Fig. 7, the p-values were associated with the Pearson's correlation between the block connectivity strength and age-regressed executive score. We note that no significant correlations were observed between the age-regressed executive function scores and covariates of no interest, including in-scanner motion and sex.

**Results**

In this study, we seek to unify the detection and quantitative characterization of modular and core-periphery structure under a single framework, and to understand how such mesoscale organization is associated with neurodevelopment and cognitive ability. To achieve this goal, we study functional brain networks estimated from resting state fMRI data acquired in 872 youth between the ages of 8 years and 22 years. Network nodes represent 333 cortical areas from the Gordon atlas (Gordon et al. 2016) (Fig. 1A) and network edges represent pairwise Pearson correlation coefficients between the regional mean BOLD time series following a Fisher's r-to-z-transformation (Fig. 1B). An increasingly common approach to studying these sorts of data is to consider *a priori* defined modules (Fig. 1C), using techniques that frequently remain agnostic to any structure of connectivity observed outside of those modules, e.g., in the off-diagonal blocks of the connectivity matrix (Fig. 1D). To provide a broader perspective, we use the weighted stochastic block model to coarse-grain the data, identifying meso-scale structure that is characteristic of each individual and that is characteristic of the group, at both finer and coarser topological scales. After characterizing this meso-scale structure, we map the association between structure and age, and reveal associations with individual differences in cognitive function. We demonstrate that a unified assessment of modular and core-periphery structure, made possible through use of the weighted stochastic block model, provides novel insights into functional network organization, its development in youth, and its implications for behavior.

**Comparison between Weighted Stochastic Block Model and Modularity Maximization**. Fundamentally, a weighted stochastic block model (WSBM) is a generative model for random graphs that seeks to partition nodes into sets such that nodes with similar patterns of weighted connectivity to the rest of the brain are grouped together with one another. The rule to group nodes by similar patterns of connectivity differs from the rules employed by other common approaches for community detection such as modularity maximization and *Infomap*. Moreover, the focus on pattern similarity allows the WSBM to simultaneously detect modules, cores, and other meso-scale structures. Specifically, if a few nodes are all densely connected to one another and sparsely connected to the rest of the brain, they will be identified as a block using the WSBM and also tend to be identified as a module using community detection techniques such as modularity maximization (Fig. 2 A). If a few nodes are densely connected to one another, and show a decrease in connectivity to the rest of brain, they will be identified as a block using the WSBM and also tend to be identified as a core using core-periphery detection techniques (Fig. 2 B). A third type of mesoscale structure that is not explicitly detected by either modularity maximization or core-periphery techniques, but which is explicitly detected by the WSBM, is bipartite structure, where one set of very sparsely intra-connected nodes is strongly and preferentially connected to another set of sparsely intra-connected nodes.

The fact that the WSBM groups nodes with similar connection patterns provides it with the flexibility to identify such diverse meso-scale structures simultaneously. To gain an intuition for the WSBM's sensitivity, we considered a group-average functional brain network constructed by taking the mean over all subject-specific connectivity matrices. To this group network, we applied both the WSBM and modularity maximization methods. We separately tuned the free parameter of both models to ensure that the subsequent solutions separated brain regions into 21 groups (a number that we will justify further below), referred to as *blocks* in the case of the WSBM and as *modules* in the case of modularity maximization. We observed that the modularity maximization approach produced a nonuniform distribution of module sizes, tending to group nearly half of the nodes into a single module (Fig. 2C). In contrast, the WSBM produced a more uniform distribution of block sizes, tending to group nodes into similarly sized blocks (Fig. 2D).

To determine the generalizability of this observation across subjects and across parameter choices, we considered the functional connectivity matrices derived for each subject separately. We then applied both modularity maximization and the WSBM to each matrix, and varied the free parameter in both algorithms to sweep across scales of the network's community architecture. We assessed performance with two metrics. First, we calculated the variance in the edge weights located within blocks or modules (Eq. 13); this metric served to probe the fit of the model to the data, with low edge weight variability indicating greater internal homogeneity of blocks or modules. Second, we calculated the variance in the sizes of blocks or modules; this metric served to probe the capacity of the method to assess the presence of community structure evenly across the network, with low community size variability indicating the capacity to detect blocks or modules of similar size. We observed that the WSBM achieved a lower standard deviation in both metrics (Fig. 2 E-F). The more evenly distributed nature of the WSBM blocks suggests that the method has the potential to discover richer meso-scale architecture.

**Hierarchical Meso-Scale Structure of Resting State fMRI in Youth.** After noting broad dissimilarities between the partitions obtained from modularity maximization and the WSBM, we next sought to better understand the full meso-scale organization of subject-level resting state connectivity matrices in the youth of the Philadelphia Neurodevelopmental Cohort. Moreover, we wished to extract features of that meso-scale organization that were specific to individuals as well as features that were conserved across the group. We therefore developed a multi-step clustering procedure, where we obtained an estimate for the block structure characteristic of single individuals, an estimate for the block structure characteristic of fine-scale organization in the group (Fig. 3A, C), and an estimate for the block structure characteristic of coarse-scale organization in the group (Fig. 3B, D).

In first considering the block structure characteristic of single individuals, we note that the WSBM has one important free parameter: k, or the number of blocks. To determine the impact of this parameter on the block solution, and to offer statistical support for a specific choice of $k$, we varied $k$ in 4,7,10,13,16,19,22,25,28 and estimated the model's goodness of fit with the log-evidence for each subject (see Eq. 2 in Methods and Fig. S1 in the Supplement). We observed that the goodness of fit first increased as $k$ increased from 4 to 16, and then appeared to plateau for $15 < k < 28$. As a second measure of reliability and robustness, we calculated the average block allegiance matrix (see Eq. 4 in Methods) over all subjects for each $k$ in the above range. We found that matrices were highly similar to one another for $k > 10$, as measured by a Pearson correlation coefficient between the vectors representing the upper triangles of the matrices (Fig. S1B). Given these tests, we chose to set $k = 16$ for each subject, which is in fact a value that is similar to that chosen for the number of functional modules or cognitive systems in previous literature (Power et al. 2007; Yeo et al. 2011).

We next turned to the question of understanding consistent block structure characteristic of all youth in the sample. We applied the WSBM to the average block allegiance matrix, and again varied the number of blocks from 3 to 30 (Fig. S1). We observed maximal log-evidence for 21 blocks (see Eq. 2 in Methods and Fig. 3A). That a larger number of blocks is required to accurately fit the group-level data in comparison to the individual-level data is expected: with the added statistical power of 872 subjects, we are able to accurately observe structure at a finer scale within the matrix. Next, we calculated the block-level average adjacency matrix as the average strength of connectivity within each of the $21 \times 21$ blocks across subjects (see Eq. 5 in Methods). By visual inspection of this matrix, we note the existence of a non-zero edge weights in off-diagonal blocks, and we also note that this nontrivial inter-block connectivity displays a heterogeneous pattern indicative of complex meso-scale architecture, which is not well-described by the simpler notion of modularity (Fig. 3C).

We note that group-level blocks differ appreciably in size (Fig. 3C) and in their spatial extent across the brain (Fig. 3A). It is therefore compelling to ask whether there exists a meaningful coarse-grained summary of these 21 blocks that accurately describes the spatial organization of the brain in broader strokes. In order to evaluate whether such a hierarchal structure exists, we applied the WSBM to the block-level average adjacency matrix (of size $21 \times 21$; see Eq. 5 in Methods), and observed maximal log-evidence for 3 blocks (Fig. S1). This coarse-grained solution segregates the functional brain network into three groups composed of regions that are reminiscent of systems defined in prior reports, including a fronto-temporal system, a default mode system, and a sensory-motor system (Gordon et al. 2016).

**Cores, Peripheries, and other Building Blocks of Mesoscale Structure.** Next, we used the WSBM to investigate how modular structure and core-periphery structure might co-exist in brain networks. Specifically, we examined each block and asked whether some blocks were more core-like while others were more periphery-like. To assess the degree to which a given block played the role of a network core, we calculated the core score of each node in the network which assesses the node's relevant associations to dense versus sparse blocks (see Methods), and then we averaged these values over nodes in a block to obtain a core score for the block (Rombach et al. 2014). Our null hypothesis was that core scores would be uniformly distributed across regions, and therefore also uniformly distributed across blocks. We first observed that core-scores were heterogeneously distributed across regions of the cortex, with highest values present in the motor strip (Fig. 4A, *top*). Across blocks, the average core-score was also heterogeneously distributed, with 6 blocks displaying greater core-scores than expected (FDR correction with $q < 0.1$, $p < 0.05$ for non-parametric permutation test in which region labels were permuted uniformly at random), and 9 blocks displaying weaker core-scores than expected (Fig. 4A, *bottom*).

We next evaluated the coarse-grained group-level structure constituting the 3 large blocks shown in Fig. 3B. For each block, we computed the core-scores of each region assigned to that block, and we examined the distribution of core-scores within a block. In the block reminiscent of the default mode network, we observed that the posterior cingulate and medial orbito-frontal areas displayed higher core-scores than expected, and that the supramarginal, middle frontal, and inferior frontal areas displayed lower core-scores than expected (FDR $q < 0.1$; $p < 0.05$; Fig. 4B, *top*). Across blocks, the average core-score was also heterogeneously distributed, with 1 block displaying a greater core-score than expected (FDR $q < 0.1$, $p < 0.05$), and 1 block displaying a weaker core-score than expected (Fig. 4B, *bottom*). In the block consisting predominantly of sensory regions, we observed a pattern of core-scores that is consistent with that observed in the whole brain, with highest values in the right temporo-parietal junction (FDR $q < 0.1$; $p < 0.05$; Fig. 4C, *top*). Across blocks, the average core-score was also heterogeneously distributed, with 4 blocks displaying greater core-scores than expected (FDR $q < 0.1$, $p < 0.05$), and 5 block displaying weaker core-scores than expected (Fig. 4C, *bottom*). In the block consisting predominantly of fronto-temporal regions, a few spatially distributed areas displayed higher core-scores than expected, but no blocks were significantly different from the null model (FDR $q < 0.1$; $p < 0.05$; Fig. 4D, *top*).

Collectively, these results indicate that the WSBM identifies blocks that play variable roles within a global core-periphery structure, and therefore motivates a more thorough examination of the nature of those roles.

**Interactions Between Blocks in the Meso-Scale Structure.** We sought to better understand how blocks interact with one another, and whether we could distinguish important principles guiding inter-block connectivity. We began by considering a pair of blocks, which is the smallest unit in which inter-block connectivity can be studied. From the relative strength between the two blocks in a pair, we could determine whether the two blocks were relatively independent, or tended to interact in either a core-periphery or bipartite manner. In applying this heuristic (Fig. 5A, see Methods section for details) to the data, we found a preponderance of independent pairs, some core-periphery pairs, and only a few bipartite pairs. To assess statistical significance, we considered a non-parametric permutation based null model in which nodes are randomly assigned to blocks. We observed that the true data displayed a greater number of independent pairs and a smaller number of core-periphery pairs than expected in the null model (Fig. 5B), indicating that core-periphery architecture, while present, complements a broad segregation consistent with the modularity commonly studied in resting fMRI.

In the exposition that follows, we did not consider the independent pairs, as it is impossible (by definition) to infer principles of inter-block connectivity from total independence. We also neglected bipartite pairs due to their infrequent existence, hampering statistical power in hypothesis testing (0.8 per subject on average). Focusing on core-periphery pairs, we first wished to determine whether their anatomical location was consistent across subjects. To address this question, we defined a characteristic measure $\eta_{mn}^{s}$ (see Eq

9), which we will refer to as the core-periphery role, to represent the core-periphery relation between blocks m and n of subject s. The value of this measure is $+1$ if block m acts as the core in the pair, $-1$ if block m acts as the periphery in the pair, and 0 if the pair does not display a core-periphery relationship. The average core-periphery role is defined as the mean of $\eta^s_{mn}$ over subjects, and intuitively, it represents the empirical probability of a pair appearing as a core-periphery pair (see Eq. 10). Here, we confine ourselves to considering the pairs that are core-periphery pairs in at least half of the participant sample: that is, the block pair mn displays an $\eta^s_{mn}$ of +1 (or of -1) in at least 50% of subjects. We find that core-periphery pairs do not appear at random locations in each subject, but instead display a consistent anatomical distribution across subjects ($z > 40$ for Eq. 11; see Methods for a description of the null model).

In addition to observing that the anatomical distribution of core-periphery block pairs was relatively conserved across subjects, we also observed that some core-periphery pairs interacted with one another in what we term *core-periphery junctions*. We observed the existence of two core-periphery junctions: one comprised of regions in the default mode system, and one comprised of regions in the executive system. The default mode core-periphery junction consisted of 3 blocks: the periphery block -- composed of regions in the rostral anterior cingulate and frontal pole -- linked two core blocks, one of which was located in the superior and medial frontal area, and the other in the precuneus and inferior parietal area (Fig. 5D). The executive core-periphery junction consisted of 5 blocks: core blocks were connected through a periphery block in a 2-tier hierarchical core-periphery structure (Fig. 5E). More specifically, the periphery block in the superior frontal area connected two core blocks: one core block was located in the supramarginal and posterior cingulate area, and the other core block was distributed across superior parietal, *pars opercularis*, and fusiform areas. Two blocks in the superior frontal, precuneus, and rostral middle frontal areas acted as the provincial cores. The presence of these two core-periphery junctions suggests an important principle by which hubs that are located in the two extreme cores might communicate with one another via shared peripheries.

**Age Related Differences in Block Structure during Neurodevelopment.** Next, we asked whether features of core-periphery structure are associated with age. We recognize that changes in topology can occur atop changes in overall network strength, and that it is important to distinguish between the two. Thus, we first examined the association between overall network strength and age (Fig. 6A). We observed that the average magnitude of the positive edge weights increased significantly with age ($r = 0.11$, $p = 0.0017$) and that the average magnitude of the negative edge weights decreased significantly with age ($r = -0.15$, $p = 1.63 \times 10^{-5}$), after partialing out the effects of sex and motion. These observations are consistent with previous reports of growing system segregation with development (Fair et al. 2007; Satterthwaite et al. 2013). To determine whether this segregation resulted from the emergence of more core-periphery interactions, we calculated the Pearson correlation coefficient between the number of core-periphery pairs and age, after partialing out the effects of sex and motion. We observed a significant positive relationship ($r = 0.19$, $p = 2.25 \times 10^{-8}$; Fig. 6B), which remains significant after partialing out the average magnitude of the positive and negative edge weights ($r = 0.16$, $p = 1.28 \times 10^{-6}$). Next, focusing solely on the two core-periphery junctions described in the previous section, we found that the interaction strength among core blocks increased significantly with age ($r = 0.15$, $p = 4.93 \times 10^{-6}$; $r = 0.21$, $p = 1.89 \times 10^{-10}$), while the interaction strength among periphery blocks was unchanged with age ($p > 0.05$; Fig. 6C), again after partialing out the effects of sex and motion. These results suggest that neurodevelopment can be described as an enhancement of core-periphery structure, driven in part by a strengthening of network cores in higher-order cognitive systems.

To better understand age related effects on the two core-periphery junctions that we identified in higher-order association areas, we considered the elemental blocks that composed each junction. In the default mode core-periphery junction composed of two core blocks sharing a periphery, we observed that the edge weights within cores increased significantly with age ($p < 0.005$; Fig. 6D). Considering the connectivity between the default mode and executive core-periphery junctions, we observed that inter-junction edges tended to decrease in weight with age (Fig. 6E), indicating a growing segregation between the two junctions. Here we report the significant effects at different levels of stringency in statistical testing to present an even-handed account. Specifically, of the 15 inter-junction relations, 13 displayed decreasing edge weight with age at a level of $p < 0.005$ and 7 displayed decreasing edge weight with age at a level of $p < 0.0005$. In the executive core-periphery junction, we observed that the edge weights within cores tended to increase significantly with age ($p < 0.0005$; Fig. 6F). Notably, all of these trends held when we computed the correlation after partialing out the effects of sex and motion. See the SI for details on the partial correlation values and associated estimates of statistical significance. Collectively, these results support the more general conclusion that development is associated with a strengthening of network cores in higher-order cognitive systems and an increasing segregation between such systems.

**Cognitive Correlates of Meso-scale Block Structure.** As mentioned previously, the weighted stochastic block model groups similarly-connected regions together into a block, and each of these blocks can play a different role in the brain's meso-scale network organization, including the role of a core and the role of a periphery. In the previous section, we showed that the interaction strength among core blocks changed appreciably over development. Here we asked whether individual differences in cognitive performance were related to the partition of nodes into blocks identified by the WSBM. If such a relationship existed, it would be important to determine whether the relationship was specific to the WSBM, or whether it could also have been found with previously developed approaches. To address this question of specificity, we tested whether individual differences in cognitive performance were also related to the partitions of nodes into modules identified from the modularity maximization algorithm (Newman 2006). Specifically, we tested for significant correlations between inter-block (or inter-module) strength and age-regressed executive function score (see Methods). We note that the age-regressed executive function scores used here were not significantly associated with sex ($r = -0.017$, $p = 0.598$) and were also not significantly associated with mean framewise displacement ($r = 0.0022$, $p = 0.949$).

In the WSBM partition, we observed 19 intra- and inter-block strengths distributed throughout the brain that were significantly correlated with individual differences in executive function (Pearson correlation coefficients, FDR corrected for multiple comparisons at $q < 0.05$). In the modularity maximization partition, we found only 3 modules whose inter-module strength was correlated with individual differences in executive function; these modules were comprised of regions in the default mode, dorsal attention, and visual systems (See Fig. S3 for details). Given the more extensive relation between the WSBM blocks and executive function, we probed the WSBM partition further. Specifically, we observed that executive function scores were positively correlated with overall core strength in both of the core-periphery junctions, and were not strongly correlated with periphery strength (Fig. 7A, B). We also noted that the block pairs whose inter-block strengths were significantly correlated with individual differences in cognition tended to be located at the centers of the two core-periphery junctions that we identified in a previous section. The interaction strength between the cores located within junctions was positively correlated with individual differences in executive function (Fig. 7C). In contrast, the interaction strength between the cores located between junctions was negatively correlated with individual differences in executive function. In assessing the sensitivity of our findings, we note that the age-regressed cognitive scores that we used here were uncorrelated with age (by definition), and they were also uncorrelated with sex and motion, and thus we did not include these three covariates in our analysis. Broadly, the pattern of results that we uncover suggests that the extent of segregation between the default mode junction and the executive junction explains significant variance in individual differences in cognitive performance.

**Discussion**

Here, we adopted the weighted stochastic model (WSBM) to investigate the nature, development, and cognitive significance of mesoscale architecture in functional brain networks. Unlike other common approaches, the WSBM is built on a generative model of graph architecture, which is sensitive to modular structure, core-periphery structure, and other mesoscale motifs (Betzel et al. 2017, 2019). Because each block is composed of nodes with a similar pattern of connectivity to the rest of the network, the method fundamentally crystallizes inter-block interactions, facilitating a rich quantitative assessment of mesoscale graph structure. By applying the method to resting state functional brain networks estimated from 872 youth ages 8 to 22 years in the Philadelphia Neurodevelopmental Cohort, we observed that blocks vary in their topological nature, with some blocks being composed predominantly of regions that play a core role within the network, other blocks being composed predominantly of regions that play a peripheral role within the network, and still other blocks being composed of regions of both types. Core and periphery roles were not only played by single regions within blocks, but they were also played by blocks themselves in their interactions with one another. Notably, we uncovered the existence of block junctions in which different cores shared the same periphery, and demonstrated that the strength of cores increased with development. Finally, we observed that individual differences in the interaction strength between cores was correlated with individual differences in executive function, particularly among the cores that participated in core-periphery junctions. Collectively, our results offer a rich depiction of mesoscale structure in functional brain networks, and highlight the role of core-periphery structure in cognition and development.

**Meso-scale Block Structure in Functional Brain Networks.** Traditionally, meso-scale structure in resting state functional brain networks has been studied from the perspective of modularity (Betzel and Bassett 2016; Sporns and Betzel 2016), where brain regions are more strongly connected to other regions in their module than expected in an appropriately defined null model (Newman 2006). The notion that modules are important for brain function, development, and cognition, has a long intellectual history (Fodor 1983), and the recent formalization of that notion with the mathematical tools of network science has led to important insights into the role of modules in development (Gu et al. 2015; Baum et al. 2017) and aging (Meunier et al. 2009). The notion of modularity has also become important in understanding individual differences in cognitive capacity such as probed by general measures of executive function in youth (Baum et al. 2017; Chai et al. 2017) as well as more specific measures of response to training both in healthy adults and in individuals with brain injury (Arnemann et al. 2015; Gallen et al. 2016). However, common tools for community detection that are applied in the neuroimaging field focus on identifying modules with algorithms that seek independent groups of brain regions (Porter et al. 2009; Fortunato 2010;

Fortunato and Hric 2016). This assumption of module independence that is implicit or explicit in common algorithms is fundamentally at odds with our intuitions, supported by neuroscience, that some modules may control, compete, or cooperate with other modules, and that these interactions might change according to context, differ in youth and the elderly, and vary in health and disease. The assumption that modules are independent also stands at odds with recent empirical evidence underscoring the importance of the patterns of inter-module connectivity for transitions between cognitive states (Cole et al. 2014; Mattar et al. 2015), long-term changes in behavior (Bassett et al. 2015), and alterations in connectivity characteristic of adolescent development (Gu et al. 2015).

An approach that embraces the potential heterogeneous patterns of interconnectivity between modules is the WSBM, which explicitly quantifies meso-scale network architecture, thereby offering insights into how control, competition, and/or cooperation between modules can be instantiated. Rather than treating modules as independent, the WSBM explicitly seeks to capture the meso-scale topology connecting modules with one another, and can be used to understand small-world organization of modules, as well as the presence of hub modules, connector modules, and provincial modules. Prior studies applying the SBM to brain networks have compared block structure and module structure (Pavlovic et al. 2014; Rajapakse et al. 2017), and have sought to better understand the diversity of mesoscale architecture consistent across species and aligned with genetic underpinnings (Betzel et al. 2017, 2019). Here we complement these prior studies by offering a systematic block-based analysis to better understand how blocks interact with one another in the wider brain network, and how those interactions might change with development or track cognitive efficiency.

**Core-Periphery Junctions.** Using the WSBM, we found compelling evidence for blocks (sets of brain regions with similar connectivity to the rest of the brain) that were best described as (i) core blocks (being strongly connected to most blocks and weakly connected to a few blocks), (ii) periphery blocks (being weakly connected to most blocks, but strongly connected to a few blocks), or (iii) blocks that -- like modules -- were neither core-like nor periphery-like. In addition to these more commonly studied meso-scale features (modules, cores, and peripheries), we uncovered the existence of block junctions in which different cores shared the same periphery. While the exact functional role of these block junctions is unknown, it is possible that they offer a means by which cores can transiently communicate with one another. Specifically, we note that peripheries are composed of regions that -- over long periods of time -- display weak static functional connectivity. However, these regions can display transient control processes or coupling dynamics over shorter time scales. Our results are consistent with the possibility that peripheries could engage in these transient dynamics as a means of facilitating communication between cores, and the network hubs that they often contain. Notably , the two core-periphery junctions that we uncovered -- containing regions of the default mode and executive systems -- were reminiscent of the task-positive and task-negative systems commonly observed in resting state fMRI, and associated with executive function (Kelly et al. 2008; Hampson et al. 2010). Notably, the periphery blocks in the center of both junctions contained portions of the medial frontal gyrus, suggesting an important mediating role for this region in mesoscale network function of resting state brain dynamics, consistent with the region's known role in top-down control of cognitive processes (Salmi et al. 2009).

**Role of Core-Periphery Structure in Development and Cognition.** Two blocks that form a core-periphery structure are collectively referred to as a core-periphery pair. We demonstrate that core-periphery pairs increase significantly in number over the developmental period of 8 to 22 years of age. Notably, we find that this increase cannot be explained by changes in the overall strength of connectivity across the brain. Rather, it is specifically driven by an increase in the strength of core blocks, a phenomenon that also serves to increase the heterogeneity of block topology and the potential for localization of functional systems. Importantly, this finding is consistent with prior observations of increasing functional segregation of cognitive systems with age (Fair et al. 2009; Gu et al. 2015). Our data put a finer point on the notion of functional segregation, providing insight into a specific topological change -- the increase in cores -- that supports the more general observation. Based on prior theories regarding the role of core-periphery organization in brain network function (Fedorenko and Thompson-Schill 2014), we speculate that the increasing core-periphery organization in youth facilitates an emerging balance between temporally invariant processes critical for task performance, and temporally transient processes critical for adaptation and control (Bassett, Wymbs, et al. 2013). To more explicitly test this hypothesis, we examined the relationship between summary metrics of this organization and individual differences in executive function. Notably, we observed a negative correlation between executive function and the strength of block interactions in the two core-periphery junctions, suggesting that the more the two junctions were anti-correlated, the greater an individual's efficiency might be in executing complex tasks. Such enhanced anti-correlation is consistent with greater segregation between the two junctions, a finding that is conceptually consistent with prior work providing evidence that age-related improvement of executive function is mediated by increasing segregation of modules (Baum et al. 2017).

**Methodological Considerations.** There are several methodological considerations pertinent to this work. First, the broad community cohort that we study here is sampled cross-sectionally, and thus we cannot address any questions related to developmental trajectories. It would be interesting in the future to consider longitudinal samples and samples enriched for deficits in executive functioning. Second, participant motion is a well-known confound that impacts the BOLD signal. In the context of studies of development, this confound is particularly important to address, as in-scanner motion tends to decrease with age. Here we address this issue with an extensively validated preprocessing pipeline that mitigates the influence of motion artifact (Ciric et al. 2017, 2018), as well as post-processing inclusion of motion as a potential confound in all statistical analyses. Third and finally, the weighted stochastic block model that we use here is applied to the static functional connectivity matrix and is therefore unable to assess any dynamic reconfiguration of network architecture over time. It would be interesting in future to consider new methods for applying the weighted stochastic block model to time-evolving graphs (Matias and Miele 2017).

**Conclusion**

Our study offers a new perspective that complements two competing perspectives in the field: one that describes functional brain networks as composed of segregated modules, and one that describes functional brain networks as composed of hubs and a rich-club. We unify the two perspectives by employing a weighted stochastic block model, which is a model-based approach with an explicitly network-based prior that can detect groups of brain regions with similar connectivity profiles to the rest of the brain. In addition to providing a unified perspective on functional brain organization, the approach that we take offers a blueprint for other future studies in other populations tackling important questions in development, cognition, and disease.

**Acknowledgments**

This study was supported by grants from the National Institute of Mental Health (to DSB and TDS): R21MH106799, R01MH107703, R01MH113550, and RF1MH116920. DSB also acknowledges additional support from the John D. and Catherine T. MacArthur Foundation, the Alfred P. Sloan Foundation, the ISI Foundation, the Paul Allen Foundation, the Army Research Laboratory (W911NF-10-2-0022), the Army Research Office (Bassett-W911NF-14-1-0679, Grafton-W911NF-16-1-0474, DCIST-W911NF-17-2-0181), the Office of Naval Research, the National Institute of Mental Health (2-R01-DC-009209-11, R01-MH112847, R01-MH107235, R21-M MH-106799), the National Institute of Child Health and Human Development (1R01-HD086888-01), National Institute of Neurological Disorders and Stroke (R01-NS099348), and the National Science Foundation (BCS-1441502, BCS-1430087, NSF PHY-1554488 and BCS-1631550). SG acknowledges the support from NSFC-61876032. Additional support was provided by the Penn-CHOP Lifespan Brain Institute. The content is solely the responsibility of the authors and does not necessarily represent the official views of any of the funding agencies.

**Figure Captions**

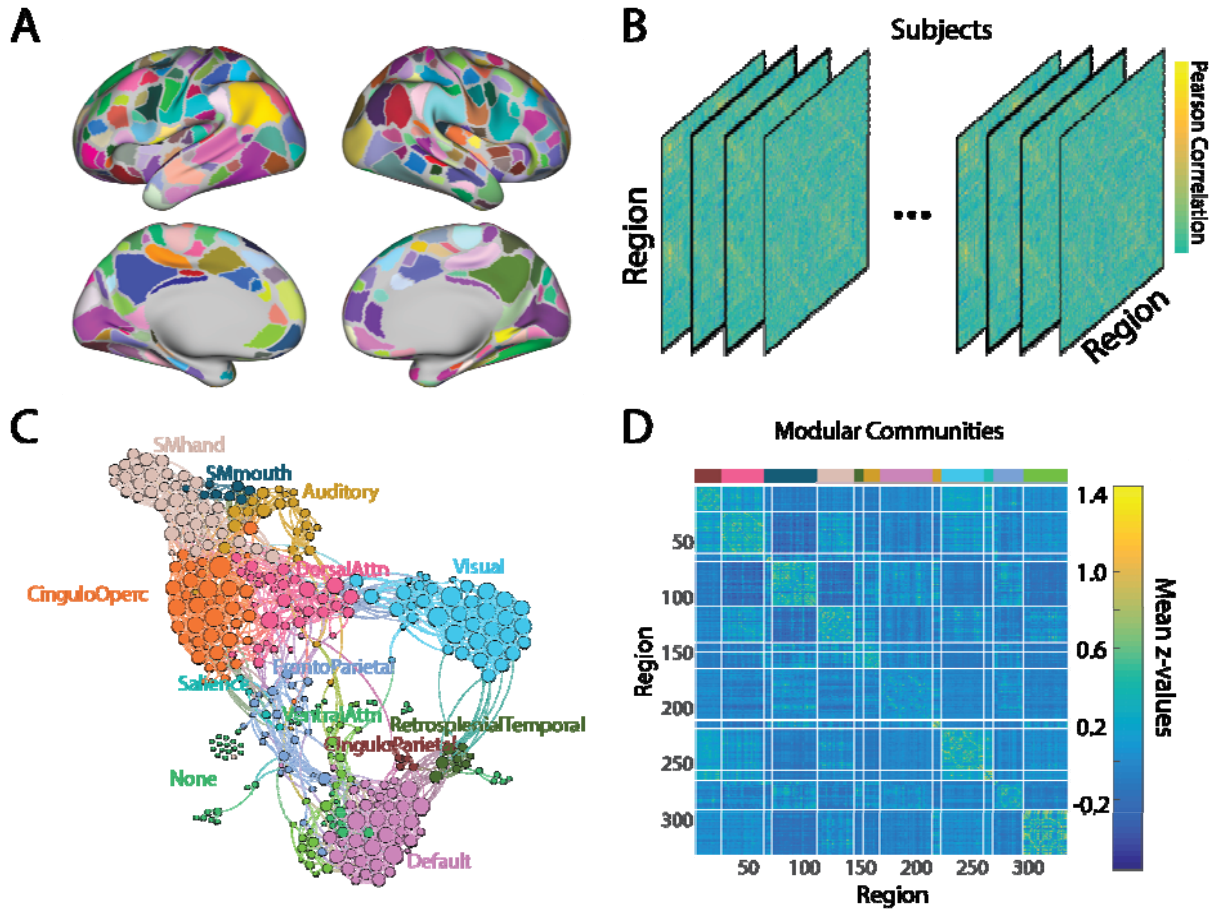

**Fig. 1 A conceptual schematic.** *(A)* We first subdivide the brain into 333 cortical areas from the Gordon atlas (Gordon et al. 2016) and extract the regional mean BOLD time series. *(B)* Next, we estimate the functional connectivity between all pairs of regions by calculating the Pearson correlation coefficient between regional time series, and applying a Fisher r-to-z-transform. For each subject, this approach produces a graph or network that we represent in an $|\mathcal{V}| \times |\mathcal{V}|$ weighted adjacency matrix. *(C)* Traditionally, this sort of network has been subdivided into functional modules based on various community detection methods. Here we show one example partition of network nodes (brain areas) into 13 modules (Power et al. 2007) forming auditory, cingulo-opercular (CinguloOperc), cingulo-parietal (CingulParietal), default mode (Default), dorsal attention (DorsalAttn), salience, fronto-parietal, retrosplenial temporal, somatomotor hand (SMhand), somatomotor mouth (SMmouth), ventral attention

(VentralAttn), and visual systems. For clarity of visualization, nodes are color coded according to their module assignment, and we show the strongest 3% of edges in the group-averaged functional brain network. *(D)* When we plot the group-averaged functional connectivity matrix with nodes ordered by the 13 *a priori* defined modules, we observe a non-zero mean and heterogeneous pattern of inter-module connectivity.

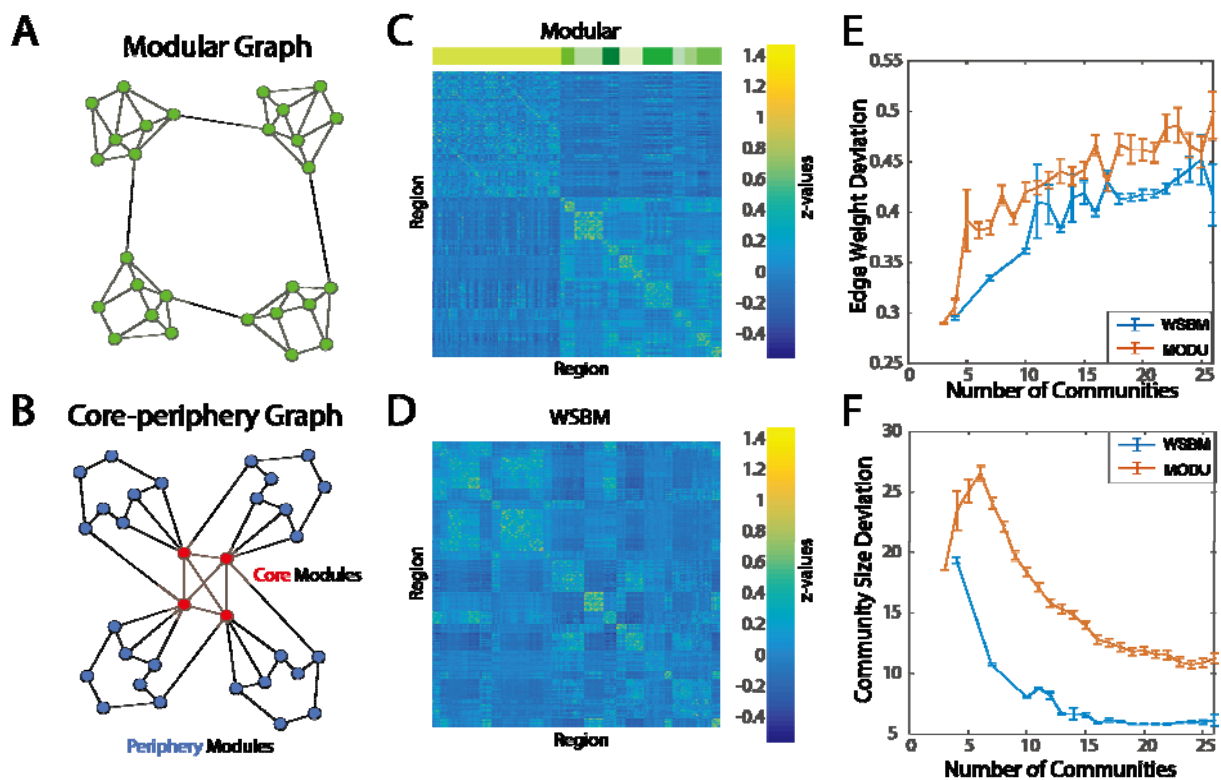

**Fig. 2 Comparison between the weighted stochastic block model and modularity maximization.**

*(A)* Modularity maximization is designed for community detection in a modular graph where the within-community connectivity is much stronger than the between-community connectivity. *(B)* Accurate and reliable detection of core-periphery structure requires a distinct methodological approach. Notably, characterizations of networks that display both modular and core-periphery structure require yet another distinct method, an example being the WSBM. *(C)* Group-averaged functional connectivity matrix with nodes ordered according to an example partition obtained from the modularity maximization approach, with the resolution parameter $\gamma$ tuned to obtain 21 modules. *(D)* Group-averaged functional connectivity matrix with nodes ordered according to an example partition obtained from the WSBM, with $k$ tuned to obtain 21 blocks. *(E)* Compared with modularity maximization approach, the WSBM recognized communities with a lower standard deviation in the edge weights located within blocks or modules (Eq. 13). *(F)* Compared with modularity maximization approach, the WSBM also recognized communities with a lower standard deviation in community size. In panels *(E)* and *(F)*, the thick line indicates the mean calculated over the 872 subjects, and the error bars indicate the standard error of the mean.

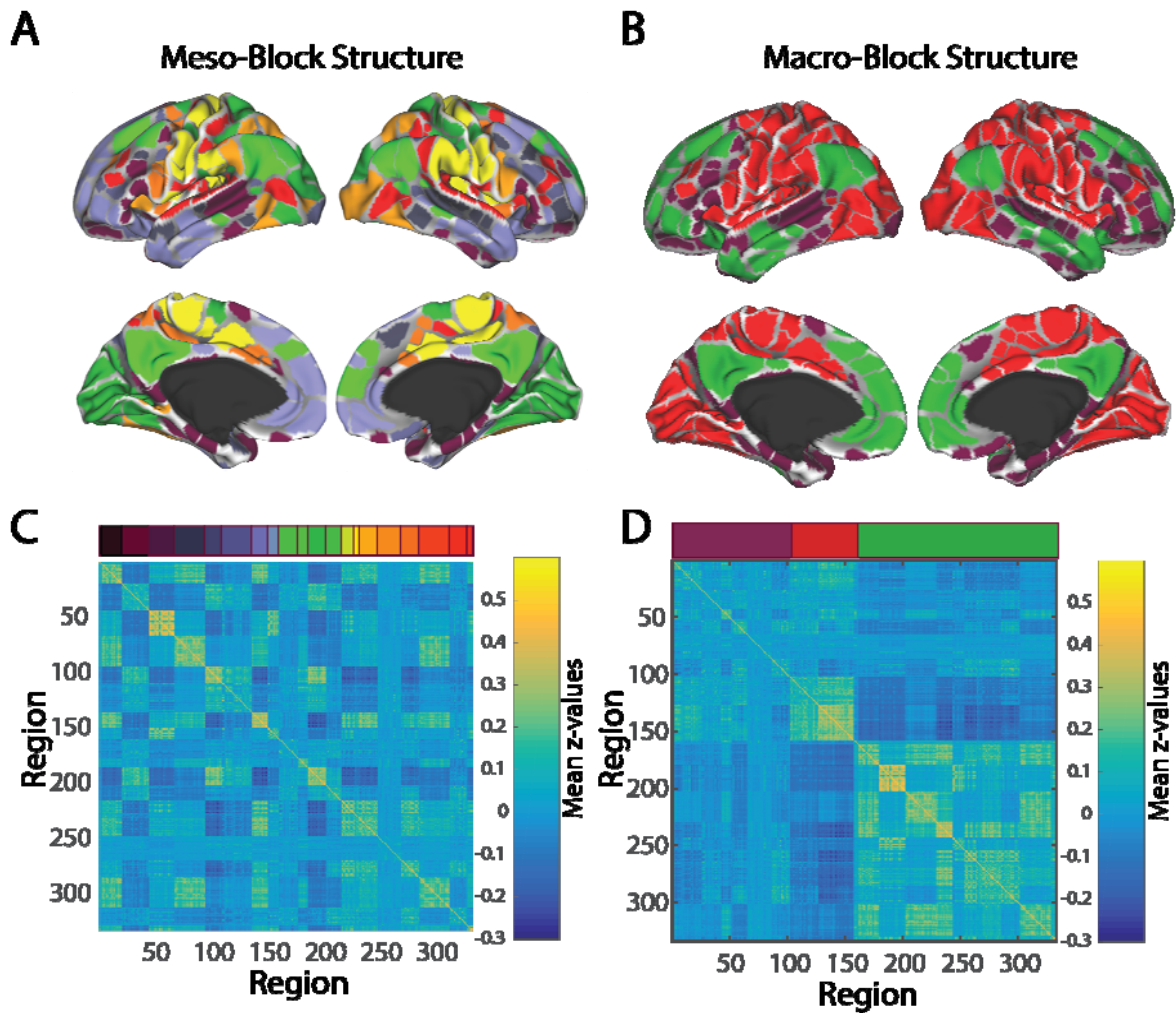

**Fig. 3 Hierarchical representation of mesoscale structure in resting state fMRI in youth.** We apply the WSBM to each functional network extracted from a single individual. *(A)* Next, we calculated the average

block allegiance matrix (see Eq. 4 in Methods) over all subjects, and applied the WSBM to this matrix to obtain the fine-scale block structure characteristic of the group. This fine-scale block structure segregates the functional network into 21 blocks. *(B)* Next, we calculated the block-level average adjacency matrix as the average strength of connectivity within each block across subjects (see Eq. 5 in Methods). Upon this average matrix we again applied the WSBM to obtain the coarse-scale block structure characteristic of the group. This coarse-scale block structure segregates the functional network into three components: a set of regions reminiscent of the fronto-temporal system, a set of regions reminiscent of the default mode system, and a set of regions reminiscent of the sensory system. Panels *(C)* and *(D)* show the reordered matrices corresponding to panel *(A)* and *(B)*, where the regions below the same color strip are located within the same block.

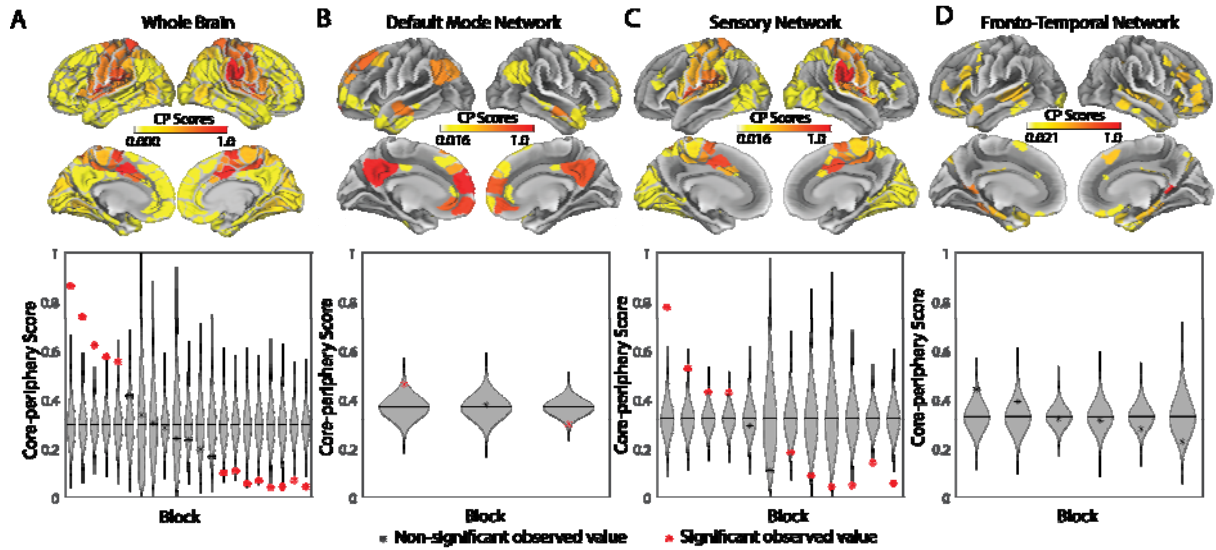

**Fig. 4 Anatomical distribution and statistical testing of core-scores.** We compute core-scores for each region both in the network representing the whole brain and in the subnetworks representing the three macro-scale blocks identified in the group-level WSBM analysis. *(A) Top* In the whole brain, the motor strip generally and the temporo-parietal junction specifically exhibited higher core-scores than expected in the non-parametric null model. *Bottom* Across blocks, the average core-score is heterogeneously distributed, with 6 blocks showing greater core-scores than expected, and 9 blocks showing lower core-scores than expected. *(B) Top* In the subgraph constituting the coarse-scale block that is reminiscent of the default mode network, we observed strongest core-scores in known default mode hubs including the posterior cingulate and ventromedial prefrontal cortex, and weakest core-scores along the lateral surfaces. *Bottom* Across meso-blocks within the default mode macro-block, the average core-score is also heterogeneously distributed, with 1 block showing a greater core-score than expected, and 1 block showing a lower core-score than expected. *(C) Top* In the subgraph constituting the coarse-scale block composed predominantly of sensory regions, we observed that the temporo-parietal junction displayed the greatest core-score. *Bottom* Across meso-blocks within the sensory macro-block, the average core-score is also heterogeneously distributed, with 4 blocks showing greater core-scores than expected, and 5 blocks showing lower core-scores than expected. *(D) Top* In the subgraph constituting the macro-block composed predominantly of frontal and temporal regions, we observed no clear anatomical localization of high core-scores. *Bottom* Across meso-blocks within the fronto-temporal macro-block, the average core-score was not significantly different than that expected in the non-parametric null model. In the bottom panels of each figure, violin plots are ordered according to the mean core-score of each block. Red asterisks indicate $p < 0.05$, FDR corrected at $q < 0.1$.

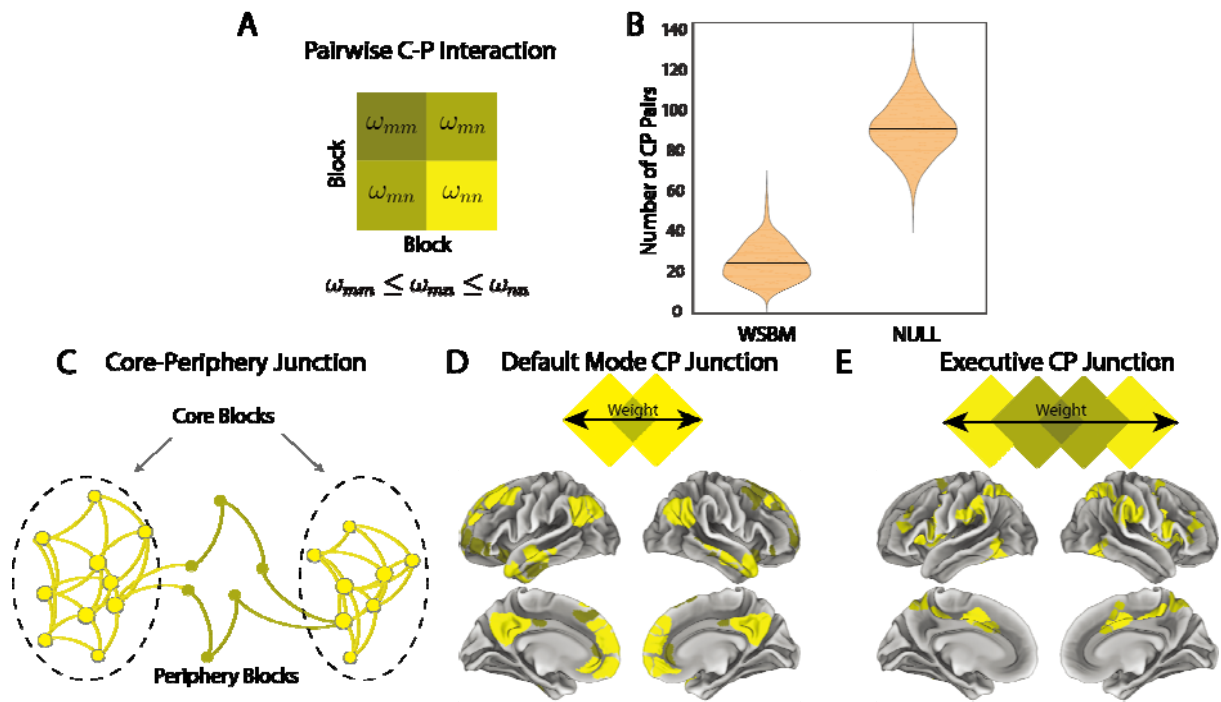

**Fig. 5 Existence of core-periphery junctions.** *(A)* For each pair of blocks, we classified an interaction as a core-periphery interaction if the average inter-block strength was intermediate between the average intra-block strength. *(B)* We compared the number of core-periphery pairs that we detected in the true data to the number of core-periphery pairs that we detected in a non-parametric permutation-based null model in which nodes are randomly assigned to blocks. We observed that the true data displayed a greater number of independent pairs and a smaller number of core-periphery pairs than expected in the null model (two-sample t-test $t = -127$, $p < 0.0001$). *(C--E)* We detected two core-periphery junctions where *(C)* different core blocks were connected through common periphery blocks. *(D)* The default mode core-periphery junction consisted of three meso-blocks: two core blocks sharing the same periphery block. *(E)* The executive core-periphery junction consisted of five meso-blocks: two hierarchical core-periphery chains sharing the same periphery block.

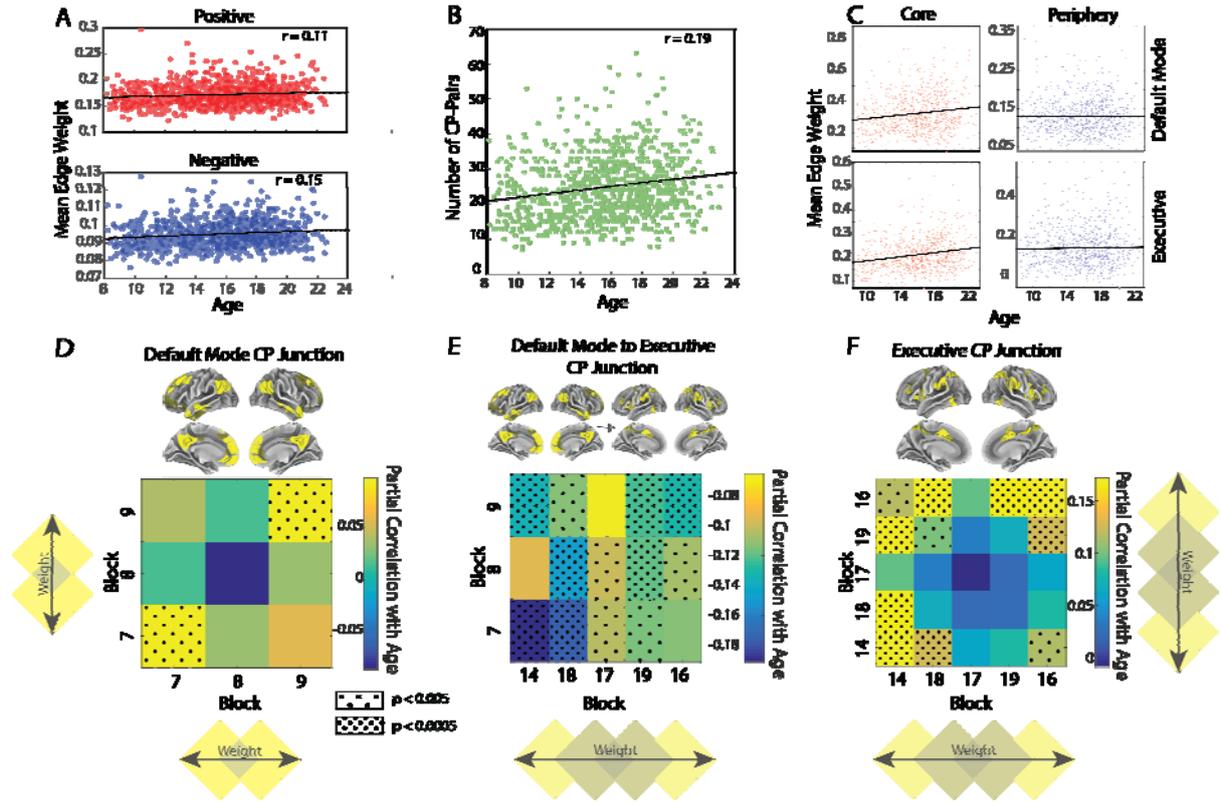

**Fig. 6 Development of core-periphery junctions in youth.** *(A)* The average strength of functional connectivity increases with age. The average magnitude of positive edge weights increases significantly with age (r = 0.11, p = 0.0017) and the average magnitude of negative edge weights increases significantly with age (r = 0.15, p = $1.63 \times 10^{-5}$). These observations are consistent with previous reports of growing system segregation with development. *(B)* To better understand changes in network topology beyond that explained by changes in the strength of connectivity, we calculated the correlation between age and the number of core-periphery block pairs. We observed a significant relation between these two variables, both based on their raw values (r = 0.19, p = $2.25 \times 10^{-8}$) and after partialing out the average magnitude of positive edge weights and the average magnitude of negative edge weights (r = 0.16, p = $1.28 \times 10^{-6}$). These results indicate an increase in heterogeneous meso-scale network architecture with age. *(C)* Focusing on the two core-periphery junctions in Fig. 5*D-E*, we next calculated the correlation between age and the average edge strength within the core and periphery blocks. We found

that the cores increase in strength over development (r = 0.15, p = 4.93 × $10^{-6}$; r = 0.21, p = 1.89 × $10^{-10}$), while the peripheries remain unchanged (p > 0.05). To further investigate the source of change in connectivity strength, we considered the *(D)* default mode, *(E)* default mode to executive, and *(F)* executive core-periphery junctions. Default mode and executive core-periphery junctions display similar changes in block strength with developmental, where the interaction strength among core areas increases with age (panels *(D)* and *(F)*). In contrast, the interaction strength between the two junctions decreases with age (panel *(E)*).

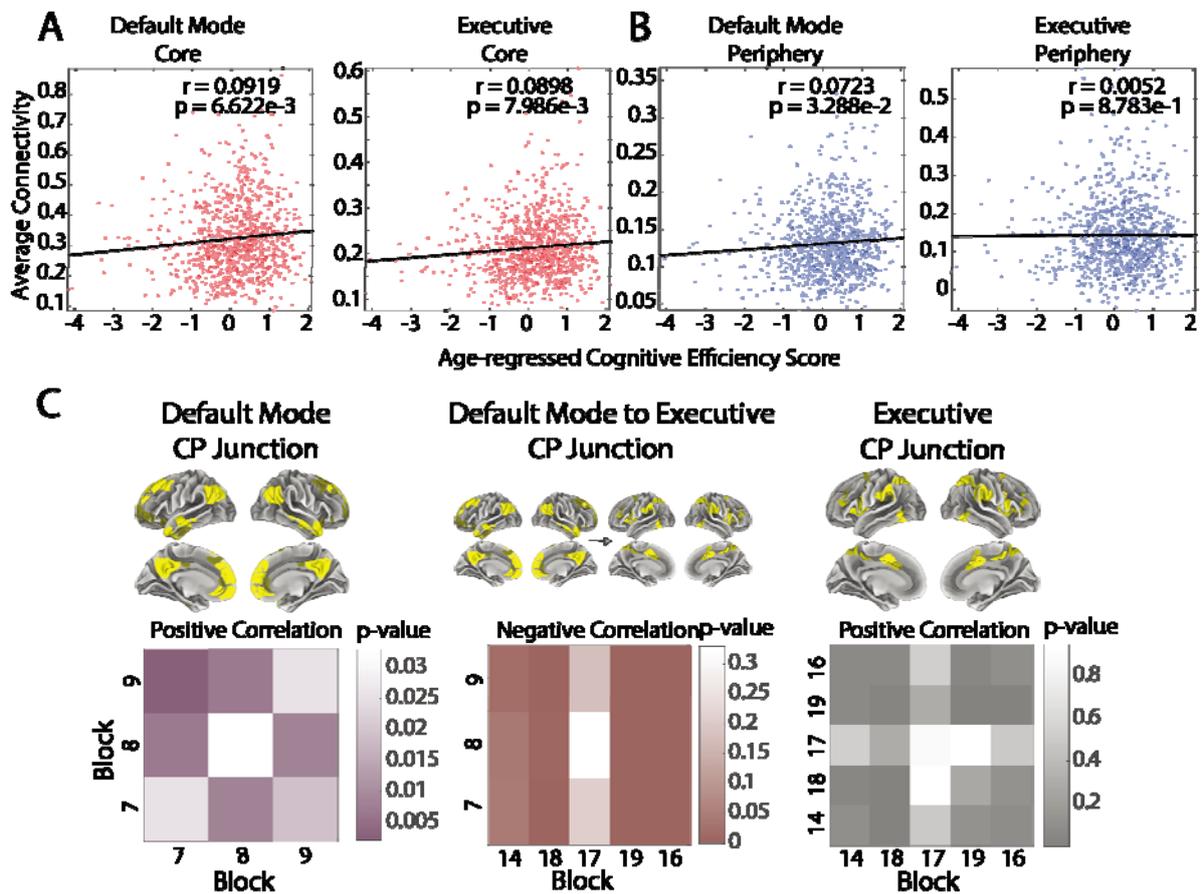

**Fig. 7 Individual differences in inter-block strength are correlated with individual differences in executive function.** For each subject, we calculated the average strength both within and between blocks

or modules. We then estimated the correlation coefficient between that value and age-regressed executive function scores across subjects. Focusing on the block structure, we next computed the correlation between the executive function scores and the connectivity strength within the core and periphery blocks recognized in the two junctions. *(A)* Overall, we find that the core strength is positively correlated with the executive function scores. *(B)* We do not find a strong correlation between periphery strength and executive function scores. *(C)* By considering the two core-periphery junctions and their interaction in greater detail, we observed that the strength of core-periphery junctions was associated with individual differences in the executive function. Specifically, we found that interaction strengths particularly among cores were significantly correlated with executive function scores, where the within-junction correlation was positive and the between-junction correlation was negative. The p-values were exploratory and not corrected for multiple comparisons.

# Supplementary Materials for "Unifying Modular and Core-Periphery Structure in Functional Brain Networks"

March 29, 2019

## Parameter Tuning for the WSBM

In the main manuscript, we used the log-evidence to measure the goodness of fit of the WSBM, and we tuned the number of blocks according to the log-evidence of the resultant block structure. We observed that the goodness of fit increased as $k$ increased from 4 to 15, but appeared to plateau for $15 < k < 28$ (Fig. 1 A). As a second measure of reliability and robustness, we calculated the average block allegiance matrices over all subjects for each $k$ in the above range. We found that these matrices were extremely similar to each other for $k > 10$, where similarity was assessed by calculating the correlation coefficient between the vectors reshaped from their upper triangles (Fig. 1 B). Based on these two points, we fixed $k = 16$ at the individual level. Next, we applied the WSBM to the block allegiance matrix to obtain the optimal group level block structure, which resulted in the 21 blocks shown in Fig. 1 C. Finally, we computed the within- and between-block average strength and constructed the $21 \times 21$ block adjacency matrix, to which we applied the WSBM for the final macro-scale block structure shown in Fig. 1 D.

## Core-periphery Junctions

In the main manuscript, we showed the existence of core-periphery junctions. Here we provide a description of the analysis steps taken to discover these junctions. From Fig. 2 A, we can see that the blocks tend to jointly constitute high level patterns. When we represented the periphery-to-core relationship as arrows, there appeared two obvious group of blocks: the default mode junction and the executive junction.

## Role of Core-Periphery Structure in Cognition

In the main manuscript, we showed that the core-periphery structure of resting state functional brain networks displayed significant correlation with executive efficiency. Here we provide a comparison with the modularity maximization approach. For the WSBM analysis, we adopted the optimal WSBM partition, which separated the resting state functional brain network into 21 blocks. For the modularity maximization analysis, we tune the resolution *gamma* so that it resulted in a similar number (finally 19) of modules. We ignored the modules with less than 4 nodes. After computing the FDR corrected $p$-values associated with the correlation between within- and between- block/module strength and the age-regressed executive function scores, we found 19 intra- and inter-block strengths displaying a significant relationship for the WSBM approach (Fig. 3A) and only 3 for the modularity maximization approach (Fig. 3B).

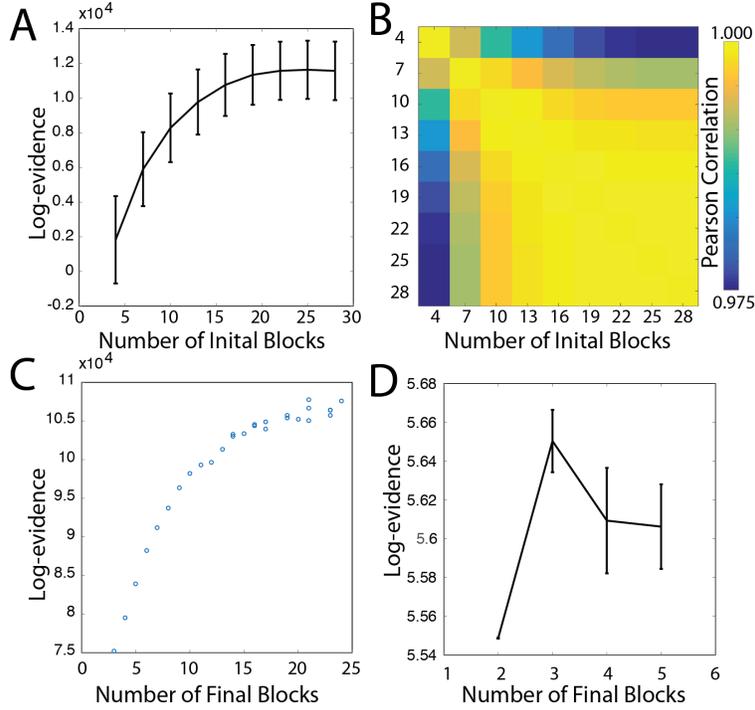

Figure 1: **Parameter Tuning for WSBM.** *(A)* We first applied the WSBM at the individual level. We found that the log-evidence increased as the number of blocks increased from 4 to 15 and it appeared to plateau for $15 < k < 28$. The error bar here shows the standard error across subjects. *(B)* We next computed the block allegiance matrix for each $k$, and we assessed their similarity by calculating the correlation coefficient between the vectors formed by their upper triangles. When $k > 10$, the pairwise correlation is over 0.99. *(C)* We fix $k = 16$ for each subject and further applied WSBM to the average block allegiance matrix to find the optimal block structure at $k = 21$, judging from the criterion of log evidence. *(D)* A final step of WSBM was performed on the block adjacency matrix to obtain the macro structure, where the optimal number of blocks is 3.

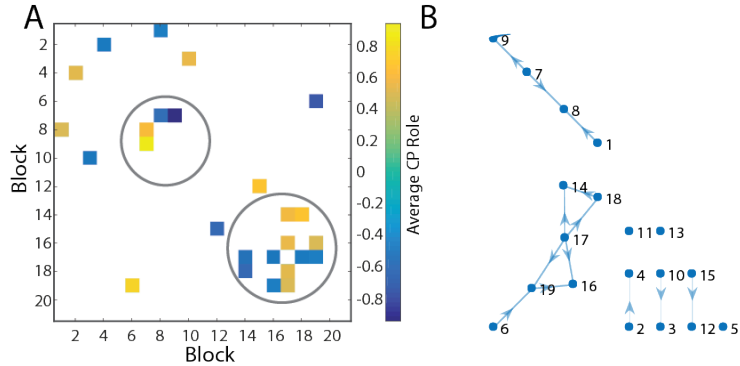

Figure 2: **Existence of core-periphery junctions.** *(A)* Instead of being isolated, several core-periphery pairs were joined at specific locations, thus forming core-periphery junctions. *(B)* If block $j$ is periphery to block $i$, we plotted an arrow from $j$ to $i$. Here we considered two typical core-periphery junctions. One contained blocks 7, 8, and 9 and the other contained blocks 14, 16, 17, 18, and 19.

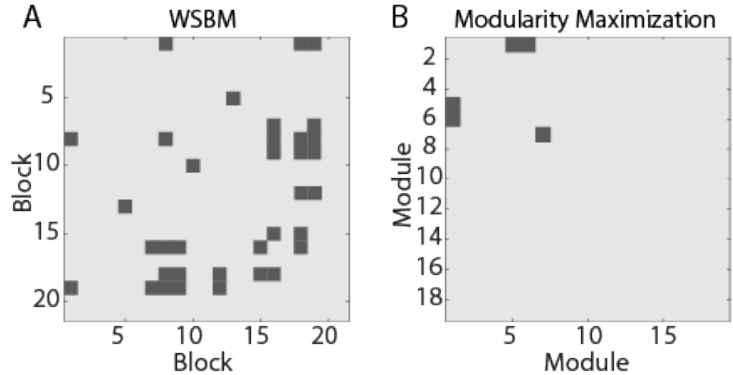

Figure 3: **Individual differences in inter-block strength are correlated with individual differences in executive function.** For each subject, we calculated the average strength both within and between blocks or modules. We then estimated the correlation coefficient between that value and age-regressed executive function scores across subjects. *(A)* In the WSBM partition, we observed 19 intra- and inter-block strengths distributed throughout the brain that were significantly correlated with individual differences in cognitive performance (Pearson correlation coefficients, FDR corrected for multiple comparisons at $q < 0.05$. *(B)* In the modularity maximization partition, only 3 module pairs displayed an inter-module strength that was correlated with individual differences in executive function.

# Robustness of Results to Sex and Motion

In the main manuscript, we showed in Fig. 6 that the block structure changes during development. Here, we provide supplementary results demonstrating that these effects are not caused by motion or by sex. The $r$- and $p$-values reported below are all calculated after partialling out sex and movement. First, we observed that the average strength of functional connectivity increases with age. The average magnitude of positive edge weights increases significantly with age ($r = 0.081$, $p = 0.016$) and the average magnitude of negative edge weights increases significantly with age ($r = 0.11$, $p = 0.0014$). Further, we observed a significant relation between age and the number of core-periphery pairs after partialing out the average magnitude of positive edge weights and the average magnitude of negative edge weights ($r = 0.17$, $p = 4.70 \times 10^{-7}$). Focusing on the two core-periphery junctions, we next calculated the correlation between age and the average edge strength within the core and periphery blocks. We found that the cores increase in strength over development ($r = 0.13$, $p = 1.06 \times 10^{-4}$; $r = 0.20$, $p = 1.50 \times 10^{-9}$), while the peripheries remain unchanged ($p > 0.05$). All of these results are consistent with those reported in the main manuscript, suggesting that neither sex nor motion can explain our findings.

# Optimal Block Structure

In this work, we reclassified the 333 regions in Gordon's parcellation into 21 small blocks and further into 3 large modules. Here we list the block label for the 333 regions and blocks in Table 1. The sensory network consisted of the blocks from 1 to 12. The default mode network contained the blocks from 13 to 16 and the fronto-temporal network contained the blocks from 17 to 21. The block IDs here were reordered by the labels of blocks in the macro structure for ease of visualization.

Table 1: **Optimal Block Structure**. This table shows the optimal block structure at the group level with 21 blocks for the 333 regions from Gordon's parcellation.

| ParcelID | Hem | Centroid (MNI) | Community in Gordon | Block | ParcelID | Hem | Centroid (MNI) | Community in Gordon | Block |
|---|---|---|---|---|---|---|---|---|---|
| 1 | L | -11.2 -52.4 36.5 | Default | 9 | 56 | L | -35.2 -35.3 42 | SMhand | 11 |
| 2 | L | -18.8 -48.7 65 | SMhand | 11 | 57 | L | -27.5 -37.2 61.4 | SMhand | 20 |
| 3 | L | -51.8 -7.8 38.5 | SMmouth | 13 | 58 | L | -47.2 -31.4 54.8 | SMhand | 20 |
| 4 | L | -11.7 26.7 57 | Default | 8 | 59 | L | -46.1 -17.8 52.7 | SMmouth | 13 |
| 5 | L | -18.4 -85.5 21.6 | Visual | 12 | 60 | L | -44.8 -54 14.6 | VentralAttn | 5 |
| 6 | L | -47.2 -58 30.8 | Default | 9 | 61 | L | -51.6 -55.9 11.4 | VentralAttn | 10 |
| 7 | L | -38.1 48.8 10.5 | FrontoParietal | 6 | 62 | L | -48.1 -40 2.4 | VentralAttn | 1 |
| 8 | L | -16.8 -60.1 -5.4 | Visual | 12 | 63 | L | -57.7 -40.6 35.8 | CinguloOperc | 16 |
| 9 | L | -55.9 -47.7 -9.3 | FrontoParietal | 4 | 64 | L | -46.3 -41.4 25.9 | Auditory | 11 |
| 10 | L | -32 -29.3 15.6 | Auditory | 13 | 65 | L | -35.8 -33.5 19.9 | Auditory | 13 |
| 11 | L | -29.3 5.3 -27.4 | None | 5 | 66 | L | -52.7 -20.6 5.4 | Auditory | 20 |
| 12 | L | -6.1 -26 28.5 | CinguloParietal | 2 | 67 | L | -59.6 -38.5 16.5 | Auditory | 20 |
| 13 | L | -14.4 -57.8 18.4 | RetrosplenialTemporal | 1 | 68 | L | -58.7 -29.9 11.1 | Auditory | 20 |
| 14 | L | -8.8 -49.8 4.2 | RetrosplenialTemporal | 1 | 69 | L | -40.6 -38.3 14.5 | Auditory | 20 |
| 15 | L | -11.3 -83.2 3.9 | Visual | 12 | 70 | L | -33.7 -21.8 9.9 | Auditory | 20 |
| 16 | L | -22 -58.1 1.5 | Visual | 12 | 71 | L | -38.7 -16 -5.3 | CinguloOperc | 14 |
| 17 | L | -9.6 -58 3 | Visual | 15 | 72 | L | -39.1 -1.6 -12.2 | CinguloOperc | 11 |
| 18 | L | -29 -35.9 -8.3 | None | 1 | 73 | L | -33.6 17.2 -31.5 | None | 1 |
| 19 | L | -18.5 -39.2 -1.1 | None | 5 | 74 | L | -43.6 36.3 8.5 | DorsalAttn | 6 |
| 20 | L | -16.7 -46 -3.7 | Visual | 15 | 75 | L | -50 20.8 10.6 | VentralAttn | 1 |
| 21 | L | -16.6 -36.1 42.7 | CinguloOperc | 18 | 76 | L | -37.7 2.9 11.7 | CinguloOperc | 14 |
| 22 | L | -9.4 -0.1 42.9 | CinguloOperc | 14 | 77 | L | -37.2 -14 19.4 | Auditory | 20 |
| 23 | L | -3.8 12.1 64.6 | VentralAttn | 1 | 78 | L | -40.3 50.4 -4.8 | FrontoParietal | 7 |
| 24 | L | -5.5 29.3 44 | FrontoParietal | 7 | 79 | L | -47.2 39 -9.1 | VentralAttn | 7 |
| 25 | L | -5.6 42.2 35.1 | Default | 9 | 80 | L | -29.1 20.5 -14 | VentralAttn | 7 |
| 26 | L | -1.7 -17.7 39.1 | Default | 7 | 81 | L | -36.6 1.4 6.4 | CinguloOperc | 14 |
| 27 | L | -8.4 14.6 33.8 | CinguloOperc | 18 | 82 | L | -37.3 8.9 -0.9 | CinguloOperc | 18 |
| 28 | L | -9 25.3 27.7 | CinguloOperc | 17 | 83 | L | -32.5 17.2 -7.8 | Salience | 2 |
| 29 | L | -10 33.9 21.5 | Salience | 2 | 84 | L | -28.8 23.7 8.4 | CinguloOperc | 16 |
| 30 | L | -10.7 -47.5 60.3 | SMhand | 20 | 85 | L | -44.3 33.2 -7.2 | VentralAttn | 1 |
| 31 | L | -15.6 -33.1 66.1 | SMhand | 13 | 86 | L | -45.4 28.8 0.8 | VentralAttn | 1 |
| 32 | L | -10.9 -29.3 69.5 | SMhand | 13 | 87 | L | -20.4 -64.6 51.4 | DorsalAttn | 16 |
| 33 | L | -6.6 -20.4 74.2 | SMhand | 13 | 88 | L | -25.8 -65 32.2 | DorsalAttn | 19 |
| 34 | L | -8 -8.7 62.9 | CinguloOperc | 11 | 89 | L | -12.7 -64.9 31.8 | CinguloParietal | 4 |
| 35 | L | -10.8 -41.1 64.9 | SMhand | 13 | 90 | L | -13.7 -77.4 26.6 | Visual | 12 |
| 36 | L | -5 -28.2 60.4 | SMhand | 13 | 91 | L | -9.9 -56.9 59.8 | DorsalAttn | 18 |
| 37 | L | -5.4 -15.9 48.8 | SMhand | 13 | 92 | L | -7.1 -63.7 54.9 | DorsalAttn | 19 |
| 38 | L | -35.8 -29.7 54.5 | SMhand | 13 | 93 | L | -10.9 -73.4 42.9 | CinguloParietal | 6 |
| 39 | L | -41.5 -12.5 50.4 | SMmouth | 11 | 94 | L | -39.3 -73.9 38.3 | Default | 9 |
| 40 | L | -42.1 -4.5 47.3 | CinguloOperc | 17 | 95 | L | -30 -74.1 36.1 | DorsalAttn | 6 |
| 41 | L | -27.3 -6.8 46.3 | DorsalAttn | 18 | 96 | L | -34.1 -61 42.4 | FrontoParietal | 4 |
| 42 | L | -27.3 1.9 52.9 | DorsalAttn | 6 | 97 | L | -31.3 -84.2 9 | Visual | 12 |
| 43 | L | -19.8 6.4 55.7 | DorsalAttn | 6 | 98 | L | -34.2 -86.6 -0.5 | Visual | 15 |
| 44 | L | -19.5 30.1 45.5 | Default | 9 | 99 | L | -43.4 -67.6 9.7 | Visual | 21 |
| 45 | L | -36.8 -22.8 61.9 | SMhand | 13 | 100 | L | -46.2 -57.7 -7.9 | DorsalAttn | 16 |
| 46 | L | -20.5 -24.9 64.5 | SMhand | 13 | 101 | L | -59.8 -4.1 8.8 | CinguloOperc | 20 |
| 47 | L | -23.4 -13.8 64.2 | SMhand | 20 | 102 | L | -52.2 -14.1 15.2 | Auditory | 20 |
| 48 | L | -17.2 -8.6 67.9 | SMhand | 11 | 103 | L | -55.1 -32.3 23 | CinguloOperc | 14 |
| 49 | L | -21.3 -0.2 62.7 | DorsalAttn | 17 | 104 | L | -50.6 -22.4 19.2 | Auditory | 11 |
| 50 | L | -28.6 -44.7 61.7 | SMhand | 11 | 105 | L | -58.8 -23.9 31 | CinguloOperc | 14 |
| 51 | L | -31.1 -48.9 47.1 | DorsalAttn | 16 | 106 | L | -45.2 2.7 32.4 | DorsalAttn | 16 |
| 52 | L | -42.9 -45 43 | DorsalAttn | 19 | 107 | L | -34.7 5.6 34 | DorsalAttn | 6 |
| 53 | L | -51.5 -11.9 29.7 | SMmouth | 13 | 108 | L | -43 19.4 33.5 | FrontoParietal | 4 |
| 54 | L | -54.1 -21.3 40.8 | SMhand | 20 | 109 | L | -40.2 23.6 23.3 | FrontoParietal | 2 |
| 55 | L | -51.7 -30.9 39.9 | DorsalAttn | 18 | 110 | L | -37.6 38.4 17.2 | DorsalAttn | 19 |

| ParcelID | Hem | Centroid (MNI) | Community in Gordon | Block | ParcelID | Hem | Centroid (MNI) | Community in Gordon | Block |
|---|---|---|---|---|---|---|---|---|---|
| 111 | L | -51.8 -0.6 5 | CinguloOperc | 14 | 167 | R | 47.9 -42.5 41.5 | FrontoParietal | 6 |
| 112 | L | -48.6 7.5 11.1 | CinguloOperc | 18 | 168 | R | 38.1 45.9 7.7 | FrontoParietal | 6 |
| 113 | L | -41.6 8.7 22.2 | DorsalAttn | 17 | 169 | R | 22.3 -46.5 -9.9 | Visual | 15 |
| 114 | L | -27.5 53.6 0 | Default | 7 | 170 | R | 59.7 -41 -10.9 | FrontoParietal | 4 |
| 115 | L | -23.4 61 -6.8 | None | 2 | 171 | R | 33.6 -22.3 13 | Auditory | 13 |
| 116 | L | -5.9 54.8 -11.3 | Default | 8 | 172 | R | 32.5 13.6 -30.5 | None | 1 |
| 117 | L | -6.8 38.2 -9.4 | Default | 8 | 173 | R | 7.6 -27 28.4 | CinguloParietal | 2 |
| 118 | L | -31.8 2.6 -16.8 | None | 5 | 174 | R | 13.8 -54.1 10.9 | RetrosplenialTemporal | 1 |
| 119 | L | -34.7 35.6 -9.6 | None | 2 | 175 | R | 15.5 -74.1 9.4 | Visual | 12 |
| 120 | L | -22.5 32.1 -13.6 | None | 5 | 176 | R | 19.6 -45.3 -4.4 | Visual | 12 |
| 121 | L | -23.8 52.2 -12.8 | None | 2 | 177 | R | 15.6 -59.6 -5 | Visual | 12 |
| 122 | L | -17.3 46.6 -17.9 | None | 2 | 178 | R | 24.9 -35.9 -4.8 | None | 5 |
| 123 | L | -13.3 24 -16.4 | None | 5 | 179 | R | 19.4 -29.9 -9.7 | None | 1 |
| 124 | L | -8.9 45.5 -20.8 | None | 1 | 180 | R | 16.2 -33.1 43.2 | CinguloOperc | 14 |
| 125 | L | -2.5 33.8 -26.2 | None | 1 | 181 | R | 6.7 5 55.9 | CinguloOperc | 18 |
| 126 | L | -63.2 -28.7 -7.2 | Default | 9 | 182 | R | 7 25.7 47.3 | FrontoParietal | 4 |
| 127 | L | -53.1 -11.4 -16 | Default | 8 | 183 | R | 8.4 34.7 22.6 | Salience | 2 |
| 128 | L | -53.2 -13 -29.2 | None | 7 | 184 | R | 7.7 44.1 5.5 | Default | 7 |
| 129 | L | -44.6 9 -37 | None | 8 | 185 | R | 8.6 4.2 40.1 | CinguloOperc | 14 |
| 130 | L | -33.8 -33.2 -15.4 | RetrosplenialTemporal | 1 | 186 | R | 3 -19.6 37.9 | Default | 7 |
| 131 | L | -28.8 -58.8 -9.1 | Visual | 12 | 187 | R | 8.8 10.8 45.9 | CinguloOperc | 18 |
| 132 | L | -34.4 -63.9 -15.7 | Visual | 15 | 188 | R | 6 21.8 32.4 | CinguloOperc | 18 |
| 133 | L | -55.1 -39.6 -16.2 | None | 2 | 189 | R | 10.3 -57.3 58.3 | DorsalAttn | 18 |
| 134 | L | -32 -3.9 -45.2 | None | 5 | 190 | R | 16.5 -32.8 67.7 | SMhand | 13 |
| 135 | L | -38.6 -13 -26.9 | None | 5 | 191 | R | 4.8 -27.1 64.8 | SMhand | 13 |
| 136 | L | -34.3 -43.8 -21.6 | Visual | 15 | 192 | R | 16.2 0.8 67.5 | CinguloOperc | 18 |
| 137 | L | -5.4 -88 18.6 | Visual | 12 | 193 | R | 11.9 -40.7 67 | SMhand | 20 |
| 138 | L | -8.6 -77.5 -3.5 | Visual | 12 | 194 | R | 5.1 -17.1 51.6 | SMhand | 13 |
| 139 | L | -41.2 -72.1 -5.9 | Visual | 12 | 195 | R | 6.8 -8.1 50.9 | SMhand | 20 |
| 140 | L | -25.2 -97.2 -7.9 | Visual | 6 | 196 | R | 8 -6.2 63.7 | CinguloOperc | 11 |
| 141 | L | -22.6 -81.7 -11.7 | Visual | 12 | 197 | R | 42.3 -11 47.3 | SMmouth | 11 |
| 142 | L | -20.5 -12.6 -23.7 | None | 1 | 198 | R | 42.5 -2.3 47.2 | CinguloOperc | 18 |
| 143 | L | -22.5 -37.1 -15 | RetrosplenialTemporal | 1 | 199 | R | 29.2 1.9 52.4 | DorsalAttn | 19 |
| 144 | L | -22 -21.9 -17.4 | None | 1 | 200 | R | 21.9 21 46.2 | Default | 7 |
| 145 | L | -15.9 48.6 37.2 | Default | 8 | 201 | R | 38.1 -22.4 60.3 | SMhand | 13 |
| 146 | L | -19.5 56.3 27.5 | Default | 8 | 202 | R | 19.7 -25 65.2 | SMhand | 13 |
| 147 | L | -26.6 46.8 20.9 | CinguloOperc | 2 | 203 | R | 29.9 -7.8 47.4 | DorsalAttn | 18 |
| 148 | L | -21.3 63.1 1.9 | FrontoParietal | 7 | 204 | R | 12.4 -28.3 69.6 | SMhand | 13 |
| 149 | L | -28.6 50.9 10.1 | FrontoParietal | 4 | 205 | R | 29.2 -13.5 64.2 | SMhand | 20 |
| 150 | L | -6.5 54.7 18.1 | Default | 8 | 206 | R | 17 -16.9 70.9 | SMhand | 13 |
| 151 | L | -15.7 64.7 13.7 | Default | 8 | 207 | R | 20.9 -6.4 65 | SMhand | 11 |
| 152 | L | -6 44.9 6.3 | Default | 7 | 208 | R | 22.6 5.6 57.6 | DorsalAttn | 19 |
| 153 | L | -28.8 38.3 28.2 | CinguloOperc | 17 | 209 | R | 29.5 -42.5 60.4 | SMhand | 20 |

| ParcelID | Hem | Centroid (MNI) | Community in Gordon | Block | ParcelID | Hem | Centroid (MNI) | Community in Gordon | Block |
|---|---|---|---|---|---|---|---|---|---|
| 154 | L | -26.2 26.6 38.8 | Default | 7 | 210 | R | 34.2 -40.6 51.6 | SMhand | 21 |
| 155 | L | -35.7 33.1 32 | DorsalAttn | 19 | 211 | R | 38.8 -42.6 40.4 | DorsalAttn | 19 |
| 156 | L | -29.3 16.8 50.7 | Default | 7 | 212 | R | 53.9 -8.3 26.1 | SMmouth | 13 |
| 157 | L | -41.7 16.1 47.5 | Default | 9 | 213 | R | 39.6 -31.5 39.7 | SMhand | 11 |
| 158 | L | -38.7 4.8 48.4 | VentralAttn | 2 | 214 | R | 28 -34.8 63.1 | SMhand | 11 |
| 159 | L | -50.8 6.9 -20.1 | None | 3 | 215 | R | 39.2 -34.6 57.5 | SMhand | 11 |
| 160 | L | -54.4 -1.4 -0.7 | Auditory | 20 | 216 | R | 37.3 -25.9 50.9 | SMhand | 13 |
| 161 | L | -59 -18 -3 | VentralAttn | 3 | 217 | R | 48.7 -26.1 52.2 | SMhand | 11 |
| 162 | R | 12.3 -51.6 34.5 | Default | 9 | 218 | R | 47.8 -15.1 49.3 | SMmouth | 13 |
| 163 | R | 20.8 -48.2 66.1 | SMhand | 11 | 219 | R | 57.5 -40.3 34.7 | CinguloOperc | 19 |
| 164 | R | 49.6 -7.4 36.1 | SMmouth | 13 | 220 | R | 48.9 -53 28.6 | Default | 9 |
| 165 | R | 11.9 21.9 59.9 | Default | 7 | 221 | R | 57.5 -45.3 9 | VentralAttn | 5 |
| 166 | R | 22 -84.6 23.7 | Visual | 12 | 222 | R | 60.9 -38.7 1.7 | VentralAttn | 7 |
| 223 | R | 54.9 -27 29.6 | CinguloOperc | 14 | 278 | R | 4.8 65.1 -7.1 | Default | 8 |
| 224 | R | 36.4 -30.7 19.4 | Auditory | 20 | 279 | R | 7.2 48.4 -10.1 | Default | 8 |
| 225 | R | 62.5 -25.6 -5.5 | Default | 9 | 280 | R | 2.9 18.7 -23.2 | None | 5 |
| 226 | R | 57.1 -17 -2.6 | VentralAttn | 3 | 281 | R | 35.1 37.3 -8.4 | None | 7 |
| 227 | R | 53.8 -15.8 5.2 | Auditory | 20 | 282 | R | 25.4 8.9 -15.7 | None | 5 |
| 228 | R | 47.4 -39.6 13.2 | VentralAttn | 10 | 283 | R | 21.2 30.3 -15.2 | None | 6 |
| 229 | R | 45.5 -37.3 3.4 | VentralAttn | 3 | 284 | R | 21.6 51.1 -14.1 | None | 2 |
| 230 | R | 59.2 -38.6 14.6 | Auditory | 20 | 285 | R | 11.9 25.7 -24.8 | None | 5 |
| 231 | R | 48.5 -26.5 -0.1 | VentralAttn | 3 | 286 | R | 13.5 20.3 -15.2 | None | 5 |
| 232 | R | 61.7 -24 1.3 | Auditory | 3 | 287 | R | 10.9 39.1 -19.7 | None | 1 |
| 233 | R | 60 -25.2 10.2 | Auditory | 20 | 288 | R | 2.2 39 -25.6 | None | 1 |
| 234 | R | 38.8 -14.4 -5 | CinguloOperc | 14 | 289 | R | 62.3 -26.4 -16 | None | 7 |
| 235 | R | 39.7 1.2 -13.1 | CinguloOperc | 20 | 290 | R | 57.5 -7.4 -16.4 | Default | 8 |
| 236 | R | 36.8 37.8 13.1 | DorsalAttn | 19 | 291 | R | 54.7 -7.8 -26.9 | None | 8 |
| 237 | R | 52.5 23.7 10.3 | VentralAttn | 2 | 292 | R | 45.2 13.6 -30.1 | None | 8 |
| 238 | R | 36.7 5.2 12.7 | CinguloOperc | 14 | 293 | R | 31.2 -45.6 -5.8 | Visual | 15 |
| 239 | R | 38.4 -12.2 20 | Auditory | 13 | 294 | R | 34.6 -35.6 -12.3 | RetrosplenialTemporal | 5 |
| 240 | R | 42.8 48.3 -5.1 | FrontoParietal | 4 | 295 | R | 34.6 -23.9 -20.4 | RetrosplenialTemporal | 5 |
| 241 | R | 48.1 38.3 -9.2 | VentralAttn | 7 | 296 | R | 20.1 -21.4 -21.5 | None | 1 |
| 242 | R | 45.2 30.7 -5.6 | VentralAttn | 1 | 297 | R | 28 -0.4 -37.3 | None | 5 |
| 243 | R | 27.4 19.7 -14.9 | VentralAttn | 7 | 298 | R | 26.9 -69.1 -6.6 | Visual | 12 |
| 244 | R | 36.6 -10 12.4 | Auditory | 20 | 299 | R | 34.9 -44 -20 | Visual | 15 |
| 245 | R | 39.6 10.4 -1.6 | CinguloOperc | 18 | 300 | R | 36.8 7.7 -37.9 | None | 1 |
| 246 | R | 36.5 5.7 6 | CinguloOperc | 14 | 301 | R | 54.5 -9.6 -37 | None | 1 |
| 247 | R | 30.6 22.8 -4.7 | Salience | 6 | 302 | R | 56.4 -27 -19.4 | None | 2 |
| 248 | R | 33.7 22.6 3.7 | CinguloOperc | 19 | 303 | R | 31.1 2.2 -46.1 | None | 5 |
| 249 | R | 34 24.4 10 | CinguloOperc | 16 | 304 | R | 39.5 -11.9 -29.7 | None | 5 |
| 250 | R | 48.1 38.4 2.4 | DorsalAttn | 19 | 305 | R | 31.6 -9.4 -35.5 | None | 5 |
| 251 | R | 26.8 -55 54.2 | Visual | 16 | 306 | R | 43.4 -24.1 -20.8 | None | 5 |
| 252 | R | 23 -66.4 51.8 | DorsalAttn | 16 | 307 | R | 13.8 -92.3 14.7 | Visual | 12 |
| 253 | R | 32.3 -63.6 33.8 | DorsalAttn | 19 | 308 | R | 10.5 -73.8 -1.5 | Visual | 12 |
| 254 | R | 15.6 -69.5 39.6 | CinguloParietal | 6 | 309 | R | 20.4 -87.3 -6.6 | Visual | 15 |
| 255 | R | 17.6 -78.3 34 | Visual | 12 | 310 | R | 5.1 -80.2 23.1 | Visual | 12 |
| 256 | R | 7.7 -85.6 31.6 | Visual | 12 | 311 | R | 14.6 -70.3 23.3 | Visual | 12 |
| 257 | R | 7.4 -69.3 49.9 | Default | 6 | 312 | R | 19.5 -10.8 -24.9 | None | 1 |
| 258 | R | 35.4 -77.1 21.1 | Visual | 12 | 313 | R | 24.5 -36.2 -13.2 | RetrosplenialTemporal | 1 |
| 259 | R | 46.5 -67.3 36.2 | Default | 9 | 314 | R | 30.4 -18.8 -19.4 | None | 1 |
| 260 | R | 41.5 -53.5 44 | FrontoParietal | 4 | 315 | R | 21 32.8 42.1 | Default | 9 |
| 261 | R | 35.7 -56.7 45.2 | FrontoParietal | 4 | 316 | R | 21.4 42.8 35.1 | Default | 7 |
| 262 | R | 33.5 -48.2 49.4 | DorsalAttn | 19 | 317 | R | 24.4 50.8 24.3 | CinguloOperc | 2 |
| 263 | R | 31.7 -85.7 2.4 | Visual | 15 | 318 | R | 31.3 39.7 25.6 | CinguloOperc | 19 |
| 264 | R | 43.8 -67.2 2 | Visual | 21 | 319 | R | 23.5 59.1 4.9 | FrontoParietal | 7 |
| 265 | R | 47.3 -52.4 -11.7 | Visual | 16 | 320 | R | 30.9 52.2 9.9 | FrontoParietal | 4 |
| 266 | R | 57 -53.8 -1.1 | DorsalAttn | 19 | 321 | R | 16 61 19.8 | Default | 9 |
| 267 | R | 49 -54.5 8.8 | Visual | 10 | 322 | R | 8.2 53.8 14 | Default | 9 |
| 268 | R | 60.9 -2.2 10.7 | Auditory | 20 | 323 | R | 5.9 54.9 29.4 | Default | 8 |
| 269 | R | 54.2 -13.6 16.9 | Auditory | 20 | 324 | R | 13.8 46.7 42.1 | Default | 9 |
| 270 | R | 53 -22.7 39.1 | SMhand | 11 | 325 | R | 6.8 44.5 34.8 | Default | 9 |
| 271 | R | 47.3 2 37.6 | DorsalAttn | 18 | 326 | R | 30.6 18.9 48.7 | Default | 7 |
| 272 | R | 37.8 28.7 35.6 | FrontoParietal | 4 | 327 | R | 42.4 19.5 48.2 | FrontoParietal | 7 |
| 273 | R | 41.8 29.1 21.6 | FrontoParietal | 6 | 328 | R | 38.9 9.6 42.7 | FrontoParietal | 4 |
| 274 | R | 50.1 3 3.9 | CinguloOperc | 14 | 329 | R | 39.7 -22.5 2.6 | Auditory | 10 |
| 275 | R | 46.6 7.8 19.3 | DorsalAttn | 16 | 330 | R | 55.8 2 -2 | Auditory | 20 |
| 276 | R | 38.6 18.8 25.5 | FrontoParietal | 6 | 331 | R | 54.4 1.1 -12.9 | Default | 3 |
| 277 | R | 28.4 57 -5.1 | FrontoParietal | 4 | 332 | R | 57.1 -6.3 -7.7 | VentralAttn | 3 |
| 333 | R | 46.6 -21.5 -8.5 | VentralAttn | 3 | | | | | |